\documentclass[12pt]{article}

\newcommand{\nn}{\nonumber}
\usepackage{graphicx}
\usepackage{cite}

\usepackage{amsmath,amssymb}

\DeclareMathOperator{\Tr}{Tr}

\newcommand{\pd}{{\partial}}
\newcommand{\ket}[1]{| #1 \rangle}
\newcommand{\cL}{{\mathcal{L}}}
\newcommand{\cA}{{\mathcal{A}}}

\newcommand{\be}{\begin{equation}}
\newcommand{\ee}{\end{equation}}
\newcommand{\ba}{\begin{eqnarray}}
\newcommand{\ea}{\end{eqnarray}}
\newcommand{\rrangle}{\rangle\!\rangle}
\newcommand{\cW}{{\mathcal{W}}}
\newcommand{\cH}{{\mathcal{H}}}
\newcommand{\cO}{{\mathcal{O}}}
\newcommand{\cQ}{{\mathcal{Q}}}
\newcommand{\cC}{{\mathcal{C}}}

\newcommand{\bA}{{\bar A}}
\newcommand{\bZ}{{\bar Z}}
\newcommand{\hW}{{\widehat{\mathcal{W}}}}
\newcommand{\no}[1]{{:\!#1\!:}}

\begin{document}

\begin{titlepage}
\thispagestyle{empty}
\begin{flushleft}
UT-06-13\hfill July, 2006 \\
\end{flushleft}

\vskip 1.5 cm
\bigskip
\renewcommand{\thefootnote}{\fnsymbol{footnote}}

\begin{center}
\noindent{\LARGE Symmetry and Integrability of}
\\
{\LARGE Non-Singlet Sectors in Matrix Quantum Mechanics}

\vskip 2cm
{\large
Yasuyuki Hatsuda\footnote{e-mail address: hatsuda@hep-th.phys.s.u-tokyo.ac.jp}
and
Yutaka~Matsuo\footnote{e-mail address:
 matsuo@phys.s.u-tokyo.ac.jp}
}
\\
{\it
Department of Physics, Faculty of Science, University of Tokyo \\
Hongo 7-3-1, Bunkyo-ku, Tokyo 113-0033, Japan\\
\noindent{ \smallskip }\\
}
\vskip 20mm
\bigskip
\end{center}

\begin{abstract}
We study the non-singlet sectors of matrix quantum
mechanics (MQM) through an operator algebra
which generates  the spectrum. The algebra is a
nonlinear extension of the $\cW_\infty$ algebra where
the nonlinearity comes  from the angular part of the matrix
which can not be neglected in the non-singlet sector.
The algebra contains an infinite set of commuting generators which 
can be regarded as the conserved currents of MQM.
We derive the spectrum and the eigenfunctions of these conserved
quantities by a group theoretical method.
An interesting feature of the spectrum  of these charges
in the non-singlet sectors is that they are
identical to those of the singlet sector except for the multiplicities.
We also derive the explicit form of
these commuting charges in terms of the eigenvalues of the matrix and show 
that the interaction terms which are typical in  Calogero-Sutherland system
 appear.
Finally we  discuss the bosonization
and rewrite the commuting charges
in terms of a free boson together with a finite number of 
extra degrees of freedom for the non-singlet sectors.
\end{abstract}
\end{titlepage}\vfill\setcounter{footnote}{0} 
\renewcommand{\thefootnote}{\arabic{footnote}} 

\newpage

\section{Introduction}

Matrix quantum mechanics (MQM) \cite{r:MQM} is described by 
an $N\times N$ hermitian matrix $X$ 
as a dynamical degree of freedom
with the upside-down potential,
\begin{equation}
 L=\frac{1}{2}\mbox{Tr}\,\left(\dot X^2\right)+
\frac{1}{2}\mbox{Tr}\,\left(X^2\right)\,.
\end{equation}
It is well-established that this model describes
the quantum gravity coupled with $c=1$ matter field.
This  system has global $U(N)$ symmetry,
\begin{equation}
 X\rightarrow UXU^\dagger\,,
\end{equation}
and the quantum states are classified according to
the representation of the $U(N)$ symmetry.
In order to describe short strings,
only the singlet sector is relevant.
It is known that the dynamics in this sector is reduced to free fermions
moving in the upside-down potential.  The system
is trivially solvable because it is a free theory.

In order to incorporate the complete physical content of
the $c=1$ matter system such as the vortices in Kosterlitz-Thouless
phase transition,
the non-singlet sectors can not be neglected \cite{r:GK,r:BK}.
It was then revealed by Kazakov, Kostov and Kutasov 
\cite{r:K3} that 2d blackholes are described by MQM,
and the non-singlet sectors play an important role because 
the dual string theory, which is described by the sine-Liouville
theory, has vortex-antivortex (winding) interactions.
In the recent works \cite{r:Mal,r:Fid}, the non-singlet sectors 
are again shown to be essential in describing
the behavior of the long open string solution of $c=1$ matter field.

Compared with the singlet sector which is described as
a free theory, the non-singlet sector is technically
more difficult since one can not neglect interactions. 
For instance, after reducing the dynamical degree of
freedom to the eigenvalues, the Hamiltonian of the system becomes
\cite{r:OM,r:BK},
\ba
 && \frac{1}{2}
\left(-\sum_{i=1}^N \frac{\partial^2 }{\partial x_i^2}
+\sum_{i\neq j}
\frac{\rho(E_{ij})\rho(E_{ji})}{(x_i-x_j)^2}
-\sum_{i=1}^N x_i^2 \right)\psi^{(c)}(x)
=E\psi^{(c)}(x)\,,
\label{cano}
\ea
where $x_i$ is the eigenvalue of $X$, $\rho$ is the representation
of the states and $E_{ij}$ is an element of $u(N)$ with
$(E_{ij})_{kl}=\delta_{ki}\delta_{lj}$.  Calogero-type interaction
is induced when the representation is restricted to the
non-singlet representation $\rho$.
Recently the Calogero-type interaction also apears in the AdS/CFT context \cite{r:AP}.

The main purpose of this paper is to provide exact microscopic descriptions
of the non-singlet sectors.
While the system is integrable, it is not a free theory. 
It makes the rigorous treatment of the upside-down 
case rather tricky at least at this moment.
Therefore, we will focus on mathematically well-defined
system with ``upside-up'' potential,
\begin{equation}
 L=\frac{1}{2}\mbox{Tr}\,\left(\dot X^2\right)
-\frac{1}{2}\mbox{Tr}\,\left(X^2\right)\,.
\end{equation}
This is easier because it is a 
collection of $N^2$ harmonic oscillators and the Hilbert space
is generated by applying a finite number of creation operators to the
vacuum.   Since  the interaction by the restriction of the representation
remains the same, it will provide a good hint to understand
the  system with the ``upside-down'' potential.  

Our strategy in this paper is to focus on the algebra
which generates the spectrum of the system.  
As we wrote, the symmetry of
the system is $U(N)$ and the Hilbert space is classified
according to the representation of this symmetry.
There are an infinite set of operators which are invariant
under $U(N)$.  Let us introduce the creation
and annihilation operator as
\begin{equation}
A=\frac{1}{\sqrt{2}}(X+iP)\,,\quad
\bar A=\frac{1}{\sqrt{2}}(X-iP)\,,
\end{equation}
where $P=\dot X$ is the momentum associated with $X$.
They satisfy commutation relations,
\begin{align}
\left[ X_{ij}, P_{kl} \right] = i \delta_{il} \delta_{jk}\,, \quad
\left[A_{ij},\bar A_{kl}\right]=\delta_{il}\delta_{jk}\,. 
\end{align}
It is easy to see that operators of the form,
\be\label{gOpm}
\cO^{(\epsilon_1\cdots\epsilon_n)}=\mathrm{Tr}(
A^{\epsilon_1}\cdots A^{\epsilon_n})\,,\quad
(\epsilon_i=\pm, A^+:=\bar A, A^-:=A)
\ee
are invariant under $U(N)$.
The multiplication of such operators to a state
does not change the representation of the state.
Thus they can be used to generate the spectrum with a specific representation.
For the singlet sector, the algebra generated by
these operators is reduced to the $\cW_\infty$ algebra
\cite{r:Winf,r:Winfother,r:Kac,r:AFMO}
whose generators are essentially the higher derivative operators 
$\oint \bar\psi(z) z^n\partial_z^{m}\psi(z)$.
This simplification occurs because the dynamical degrees of freedom of the matrix
are reduced to those of its eigenvalues and the system becomes
free fermion system.   In particular  the ordering of the matrix multiplication
in (\ref{gOpm})  becomes irrelevant.
On the other hand, for the non-singlet sector, the off-diagonal
components become relevant and one can not change
the ordering in the above sense.  Consequently the number of
independent generators are much larger than the usual $\cW_\infty$
algebra.  It will be also shown that the algebra becomes nonlinear
and the structure of the algebra is considerably different from the
$\cW_\infty $ algebra.
We will denote this nonlinear extension as $\hW_\infty$ algebra.
In a sense, the difficulty of the non-singlet sectors
is materialized in the complication of the $\hW_\infty$ algebra.  
We will nevertheless show that the
difficulty is manageable and we can derive the complete 
sets of eigenfunctions for  the infinite set of commuting
operators in  the $\hW_\infty$ algebra.  We note that 
these operators are the conserved charges of MQM
and their existence implies the integrability of MQM.

There are a few methods to analyze the Hamiltonian dynamics
with the Calogero-type interaction (\ref{cano}).  A standard approach 
is to reduce the dynamical degree of freedom to the eigenvalues
as in (\ref{cano}).  For the analysis of  the 
relatively simpler generators in $\hW_\infty$
such as the Hamiltonian itself, their representations remain relatively compact
and we have a merit of using much fewer dynamical degree of freedom.
The second approach is to apply the 
bosonization technique where the power sums of the 
eigenvalues are identified with the free boson oscillators.  
For the singlet sector, it has a definite merit that one can
represent the free fermion directly through the boson-fermion correspondence.
Even in the presence of the Calogero type interaction,  it is still
possible to use the bosonization as discussed in  \cite{r:AMOS}.  
There appear a finite number of additional degrees of freedom
from the non-triviality of the  representations.  In \cite{r:Mal},
such degrees of freedom are physically identified 
as  the ``tips'' of the folded long open string.
By following  this reference, we will refer the additional degree of freedom
that arises from the nontriviality of the representation as  
the degree of freedom of  the tips. 
These two approaches (Calogero and bozonization) 
share a merit that it has a direct interpretation
by the conformal field theory.  On the other hand, 
in the analysis of the higher conserved quantities in  $\hW_\infty$,
the representation in terms of the eigenvalues 
is getting more and more complicated.
For a systematic study of the $\hW_\infty$ algebra, the representation
in terms of the original matrix becomes much simpler.  Indeed we obtain
the explicit forms of the eigenfunctions by this approach.
We will use a representation of generic elements in the Hilbert space
as the multi-trace operators applied to the vacuum.
Suppose we identify each trace as a loop operator, the commuting charges
of the $\hW_\infty$ algebra describe splitting and joining of these  operators.
This action resembles the interaction of 
the matrix string theory \cite{r:MatStr}
and the commuting charges can be represented as the 
action of the permutation group $S_n$.  This observation enables us to
find the exact eigenstates by applying the group theory.

We organize this paper as follows.
In \S \ref{sec:Winf}, after a brief review of the
basic material of MQM, we present some properties of
the $\hW_\infty$ algebra and construct a few of their highest
weight states in the content of MQM.  
Since $\hW_\infty$ and $U(N)$ commutes,
the highest weight states of $\hW_\infty$ can be
decomposed into the irreducible representations of
$U(N)$.  It provide an efficient way to derive the explicit form
of wave functions in each specific representation.
In \S \ref{sec:eigen}, we discuss the reduction of the algebra
in terms of eigenvalues of $\bar A$.
The off-diagonal components of $A,\bar A$ provide
extra contributions to the generators of $\hW_\infty$.
The second conserved charge, for example, has 
interaction terms which look like the Calogero-Sutherland 
interactions.  In \S \ref{sec:boson}, we rewrite the
conserved charges by the bosonization technique and
present their spectrum. We emphasize that the spectrum
has an important feature that every sector share the
same spectrum for the infinite set of charges up to
the multiplicity.  Finally in \S \ref{sec:eig}, we come back
to the analysis of the $\hW_\infty$ algebra in the matrix form.
After presenting the analogy with the matrix string theory,
we derive the analytic expression of the exact eigenstates for 
any type of the representation by using Young symmetrizer.
We also discuss the relation between the eigenstates
thus derived with those from the bosonization technique
but it is so far successful only for the part of the eigenstates.
In \S \ref{sec:summary}, we give a short summary and present
a few future issues.
In the appendix \S\ref{a:o3}, we describe an $O(3)$ harmonic oscillator system.
It gives an elementary toy model where key features of MQM can be
seen.  In particular, the role of $U(N)$ and $\hW_\infty$ is replaced
by much simpler algebras $O(3)$ and $SL(2,R)$.  It is helpful
to understand the basic strategy of this paper. In appendix \ref{a:lower},
we present the explicit forms of the eigenstates which are construced in
\S\ref{sec:eig}. It illuminates the correspondence between CFT and the 
group theoretical construction of the MQM eigentstates.

\section{Non-singlet sectors in MQM and $\hW_\infty$ algebra}\label{sec:Winf}
\subsection{Basic structure of MQM}
We first present the basis material
of MQM to fix the notation.
The Hamiltonian of the system is written as,
\begin{equation}
 \cH=\frac{1}{2}\mbox{Tr}\,\left(P^2\right)+
\frac{1}{2}\mbox{Tr}\,\left(X^2\right)=
\frac{1}{2}\mbox{Tr}\left(A\bA+\bA A\right)=
\mbox{Tr}(\bar A A)+\frac{N^2}{2}\,.
\end{equation}
The ket vacuum $|0\rangle$ (resp. the bra vacuum $\langle 0|$) is 
specified by $A_{ij}|0\rangle=0$ (resp. $\langle 0|\bar A_{ij}=0$).
In this paper we will mainly work in this creation and annihilation basis
instead of working with the coordinate ($X$) representation.
The translation between the two basis can be made by
replacing $A^\mp_{ij}\rightarrow \frac{1}{\sqrt{2}}\left(
X_{ij}\pm \frac{\partial}{\partial X_{ij}}\right)$
and $|0\rangle \rightarrow \frac{1}{(\pi)^{N^2/4}}e^{-\frac{1}{2}\mathrm{Tr}(X^2)}$.
Equivalently, it can be represented by the integral transformation,
\begin{equation}\label{trcc}
 \Psi^{(c)}(X)=\int \frac{dZd\bZ}{(2\pi)^{N^2}}
e^{-\frac{1}{2}\mathrm{Tr}X^2+\sqrt{2}\mathrm{Tr}XZ-\frac{1}{2}
\mathrm{Tr} Z^2-\mathrm{Tr}Z\bZ}\Psi(\bZ)
\end{equation}
where we represent the Fock state by the coherent 
state representation,
$\Psi(\bar Z)=\langle 0|e^{\mathrm{Tr}(\bar Z A)}|\Psi\rangle$.

The Hilbert space of MQM is constructed by applying 
the creation operators $\bar A_{ij}$ to the vacuum.
The eigenvalue of $\cH_0\equiv \cH-N^2/2$ is simply the
number of the creation operators which are applied to the vacuum\footnote{
In the following we will refer the eigenvalue of $\cH_0$
as the level of the state.},
\begin{equation}
 \cH_0 \bar A_{i_1 j_1}\cdots \bar A_{i_n j_n}|0\rangle
= n \bar A_{i_1 j_1}\cdots \bar A_{i_n j_n}|0\rangle\,.
\end{equation}
In this sense, the diagonalization of the Hamiltonian $\cH$
is trivial.
Consequently the partition function
is simply given as $Z(q)=\mbox{Tr}q^{\cH_0}=(1-q)^{-N^2}$.
Nontrivial structures appear only after we impose the
restriction on representation of $U(N)$.
We also note that the wave function in terms of the 
eigenvalues of $X$ are much more complicated. 
The complication comes in when we change the 
Fock space basis to the canonical wave function of
the eigenvalues of $X$. 

The generators of $U(N)$ algebra are written as,
\begin{eqnarray}
J_\Lambda=i\Tr \left(
\Lambda\left[X,P\right]
\right)=-\Tr \Lambda\left[A,\bar A\right]\,.
\end{eqnarray}
They satisfy commutation relations,
$
\left[J_\Lambda,\cQ_{ij}\right]= \sum_k(\Lambda_{ik}\cQ_{kj}-\cQ_{ik}\Lambda_{kj})\,,
$
for $\cQ=X,P,A,\bar A$.
Since they commute with the Hamiltonian
$\left[J_\Lambda,\cH\right]=0$,
the quantum Hilbert space can be classified according to
the irreducible representation $\rho$ with respect to $U(N)$,
\begin{equation}
J_\Lambda |\Psi\rangle_a=\sum_{b=1}^{\dim \rho}\rho(\Lambda)_{ab}|\Psi\rangle_b\,,
\;\;(a=1,\cdots,\dim \rho)\,.
\end{equation}

\begin{figure}[tbp]
	\begin{center}
		\includegraphics[width=12cm]{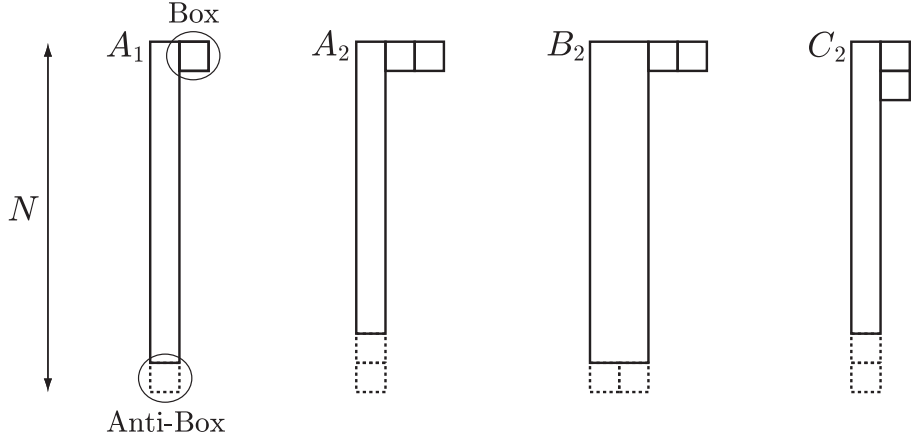}
	\end{center}
	\caption{Simpler representations allowed in MQM.}	
	\label{fig:young}		
\end{figure}

Since the wave function is constructed by combining $\bar A$
with the adjoint representation, the admissible representations $\rho$
are restricted.  The possible ones are those which correspond to
the Young diagram with the same number of boxes and anti-boxes \cite{r:BK}.
These representations are produced by 
direct products of the adjoint representations.
For example, by a direct product of two adjoint operators,
\begin{align}
	A_1 \otimes A_1 =S \oplus 2A_1 \oplus A_2 \oplus A_2^* \oplus B_2 \oplus C_2\,,
\end{align}
where $S$ is the singlet (trivial), $A_1$ is the adjoint, 
and $A_2, B_2, C_2$ (in the notation of \cite{r:BK})
are the representations with two boxes
and two anti-boxes (fig.\,\ref{fig:young}).
The simplest representation is the singlet 
$\rho(\Lambda)=0$.
The next simplest one is the adjoint,
\be
J_\Lambda|\Psi\rangle_{ij}=\sum_k (\Lambda_{ik}|\Psi\rangle_{kj}
-\Lambda_{kj}|\Psi\rangle_{ik})\,.  
\ee
We note that since the integration kernel in (\ref{trcc}) is
invariant under $U(N)$, the states in a specific representation
$\rho$ in the creation/annihilation operators is mapped to
the wave function $\Psi^{(c)}(X)$ with the same representation $\rho$.

The partition function becomes nontrivial after
the restriction of the representation%
\footnote{
We observe
in appendix \ref{a:o3} that
there is a close analogy with $O(3)$ harmonic oscillator
system.  In that case, the symmetry of the system is $O(3)$
whereas the spectrum generating algebra is given by $sl(2,R)$.
The partition function of the three harmonic oscillators
has a similar decomposition (\ref{so3tot}).
},
\be\label{decm}
Z(q)=\sum_{R}d_R Z^{(R)}(q)\,,
\ee
where the summation over $R$ is for the admissible representation
of $U(N)$, $d_R$ is the dimension of the representation
and $Z^{(R)}(q)$ is the generating function of  the multiplicity of the representation
at level $n$ as the coefficient of $q^n$.  
For simpler representations, they are explicitly written as \cite{r:BK},
\begin{align}
	Z^{(S)}(q)&=\prod_{n=1}^N \frac{1}{(1-q^n)}\,,\quad
	Z^{(R)}(q)=P^{(R)}(q)Z^{(S)}(q)\label{part1}\,,\\
	P^{(A_1)}=\frac{q-q^N}{1-q}\,,\quad
	&P^{(B_2)}=\frac{q^2(1-q^{N-1})(1-q^N)}{(1-q)(1-q^2)},\quad
 P^{(C_2)}=\frac{q^2(1-q^{N-3})(1-q^N)}{(1-q)(1-q^2)},
\nn\\
        &
	P^{(A_2)}=\frac{q^3(1-q^{N-2})(1-q^{N-1})}{(1-q)(1-q^2)}\,,\label{part2}\\
	d_{B_2}=\frac{N^2(N+3)(N-1)}{4}&,\quad
	d_{C_2}=\frac{N^2(N-3)(N+1)}{4}\,,\quad
	d_{A_2}=\frac{(N^2-1)(N^2-4)}{4}\,.\label{part3}
\end{align}

At this point, it is possible to write the strategy of 
our study more precisely.  
\begin{enumerate}
 \item We give a systematic derivation of the MQM Hilbert space 
with specific irreducible representations of $U(N)$.
Our strategy is to find the states with the specific
representation $\rho$ of $U(N)$ with the lowest $\cH_0$
eigenvalue as the highest weight state\footnote{\label{f1}
We keep the terminology of ``highest weight representation''
while the state indeed has the lowest weight. It is for keeping the convention
of  CFT.  The definition of the degree of the operator in the following has
also the different sign compared with the usual CFT convention.}
 of the $\hW_\infty$ algebra.
The generic states with the representation $\rho$ can be generated
from this set of states by applying the $\hW_\infty$  operators.
It explains the decomposition (\ref{decm}) where $Z^{(R)}(q)$
are regarded as ``characters'' of the representations of 
$\hW_\infty$.
 \item As we noted, the diagonalization of the Hamiltonian
is trivial in the creation/annihilation basis
since it just counts the number of creation operators.
These states are, however, not convenient for many purposes
since they are not diagonal with respect to the inner product.
As we will see, the $\hW_\infty$ algebra contains the infinite
set of commuting charges which include $\cH_0$ as the simplest
charge.  These operators are also important since their existence supports
the integrability of MQM in the non-singlet sectors.
We will construct the basis of the Hilbert space
which are the eigenvectors of these infinite set of charges.
\end{enumerate}

\subsection{$\hW_\infty$ algebra}

As we have defined in the introduction, 
the $\hW_\infty$ algebra %
is generated by the operators of the form
(\ref{gOpm}) which commute with $U(N)$ generators,
\begin{equation}\label{commu}
 \left[J_\Lambda,\cO^{(\epsilon_1\cdots \epsilon_n)}\right]=0\,.
\end{equation}

Before we study some detail of the algebra, it is
useful to summarize  our nomenclature on this algebra.
We first note that operators of the form (\ref{gOpm})
are eigenstates of $\cH_0$,
$
 \left[\cH_0,\cO^{(\epsilon_1\cdots \epsilon_n)}\right]
= m \cO^{(\epsilon_1\cdots \epsilon_n)}\,,
$
where $m$ is the number of $\bar A$ minus the number of $A$.
We call this eigenvalue of $\cO^{(\epsilon_1\cdots \epsilon_n)}$
as the {\em degree} of this operator.
Obviously by applying degree $n$ operator to level $m$
state produces a level $n+m$ state.
Suppose we have a set of states $\{|a\rangle\}$ at level $n$
with some representation $\rho$ under $U(N)$ {\it i.e.}
$J_\Lambda|a\rangle = \sum_{a}\rho_{ab}|b\rangle$.
The commutativity (\ref{commu}) between $J_\Lambda$
and the $\hW_\infty$ generator implies that
$\cO^{(\epsilon_1\cdots \epsilon_n)}|a\rangle$ belongs
to the same representation at level $n+m$ where
$m$ is the degree of $\cO^{(\epsilon_1\cdots \epsilon_n)}$.
Since the level is bounded from below, there must be
a set of states $|R\rrangle_a$ ($a=1,\cdots,\mathrm{dim}(R)$)
which are annihilated by all the operators in $\hW_\infty$
with negative degree.  We call such set of states
as the {\em highest weight state}  (see footnote \ref{f1}) of 
the $\hW_\infty$ algebra
for the representation $R$. 

Here we should not confuse  the highest
weight conditions for $U(N)$ and $\hW_\infty$.
The former picks up one state in each set of states $\{|a\rangle\}$
which spans basis  of the irreducible representation $\rho$.
Such states exist at various levels.  Partition functions which count
 such states are given as $Z^{(R)}(q)$. On the other hand,
the latter condition picks up the representation space
of $R$ with $d_R$ states at the lowest level.   

It is also convenient to introduce
the {\em normal ordering} prescription.
As in the quantum field theory, we define
the normal ordered operator
$\no{\cO}$ by putting all the annihilation operator $A$ on
the right of creation operators $\bar A$.  For example,
\begin{equation}
 \no{\cO^{+-+-}}=\no{\mbox{Tr}(\bar A A\bar A A)}
=\sum_{i,j,k,l}\bar A_{ij}\bar A_{kl}
A_{jk}A_{li}\,.
\end{equation}
We note that the ordering of the matrix multiplication is
not changed.  By using normal ordered operator, one
can avoid the unnecessary factors of order $N$ (and higher)
in the operator algebra.
It also makes it possible to recover the cyclicity
in the trace, $\no{\mbox{Tr}(\cO_1\cO_2)}=\no{\mbox{Tr}(\cO_2\cO_1)}$.

The algebra between the $\hW_\infty$ generators are far more
complicated than the usual $\cW_\infty$ algebra 
and it seems not possible to write the whole algebra
in a closed compact form.  The complication comes in from
the order dependence of operators and the inclusion
of higher order differential operators.
Instead of trying to write the whole algebra,
we just present the algebra between
a few simpler operators which contains  a few $A$.
It is enough to show some nontrivial features of the algebra.
We define $J_n=\mbox{Tr}(\bar A^n)$,
$L_{n}=\mbox{Tr}(\bar A^{n+1}A)$
and $V_{n,m}=\no{\mbox{Tr}({\bar A^{n+1}A\bar A^{m+1}A})}$.
The commutation relations between these operators become,
\begin{eqnarray}
&& [J_n,J_m]=0\,,\quad
 [J_n,L_m]=-n J_{n+m}\,,\quad
 [L_n,L_m]=(m-n)L_{n+m}\,,\label{com1}\\
&& \left[V_{n,m}, J_r\right]=r\sum_{s=1}^{r-1}
J_{n+s}J_{m+r-s}
+2r L_{n+m+r}\,,\label{com2}\\
&&[V_{n,m},L_r]=2\sum_{s=0}^r V_{n+s,m+r-s}
+\sum_{s=1}^{r}(r-s+1)(J_{m+s}L_{n+r-s}+J_{n+s}L_{m+r-s}) \notag \\
&&\hspace{62pt}-(m+1)V_{n,m+r}-(n+1)V_{n+r,m}\label{com3}\,.
\end{eqnarray}
The algebra in the first line is identical with
the usual $U(1)$ current algebra and Virasoro algebra
without the central extension. 
It implies that we can keep some features of CFT even for the non-singlet
sectors. The nonlinearity is a characteristic feature of the $\hW_\infty$ algebra
and it shows up in the algebra of $V_{n,m}$
in (\ref{com2},\ref{com3}).
In general, the $\hW_\infty$ generators induce 
splitting and joining of the multi-trace operators
as we will see in section \ref{sec:eig}.
The above nonlinearities are simple examples of such
property.

Finally, among the degree zero operators in $\hW_\infty$,
there are the infinite set of generators which 
commute with each other (the elements of
Cartan sualgebra). We define\footnote{
It coincides with the definition of the conserved charges in
\cite{r:Poly_rev}.},
\begin{equation}
 {H}_n=\no{\mbox{Tr }(\mathcal{L}^n)}\,,\qquad
 \mathcal{L}={\bar A A}\,,\quad
 n=1,2,3,\cdots\,.
\end{equation}
In order to prove that they commute with each other,
we use the commutation relation,
$\left[\cL_{rs},\cL_{tu}\right]=\cL_{ru}\delta_{st}-\cL_{ts}\delta_{ru}$.
Commutation relations $\left[\cL_{rs}, H_n\right]=0$ and 
$\left[H_n, H_m\right]=0$  follow from this algebra immediately.
We note that this set of commuting generators contains the
Hamiltonian $\cH_0$ as the first generator $H_1$.
We also remark that there are many other operators with degree zero in $\hW_\infty$
such as $\mbox{Tr}(\bar A^2 A^2)$.  These operators, however, do not
commute with $H_n$ and can not be taken as elements of Cartan subalgebra.

\subsection{Construction of non-singlet states by the $\hW_\infty$ algebra}
In the following, we give a few explicit constructions
of the non-singlet sectors by using the $\hW_\infty$ algebra.
As we have explained, we first
find the highest weight state of the $\hW_\infty$ algebra
which provide the non-singlet states at the lowest level.
The generic non-singlet states can be generated from
this highest weight vectors by applying the $\hW_\infty$ generators
with positive degree.
In order to identify the irreducible set of states, it is convenient
to re-order the product by using commutation relations to
the following standard form,
\begin{equation}
\prod J_s \prod L_r\prod V_{n,m}\cdots |R\rrangle\,.
\end{equation}
Namely we reorder the operators which contains more $A$ to the right side
of the operators with less $A$.
In the following, we present the highest weight states
at level 0,1,2.

\paragraph{Highest weight condition at level 0 and the singlet sector}
There is only one state (Fock vacuum $|0\rangle$)
and it automatically satisfies the
highest weight state condition.
Since $J_\Lambda|0\rangle=0$, this is the state
which belongs to the singlet representation.
Therefore, we write,
\begin{equation}
 |S\rrangle=|0\rangle\,.
\end{equation}
It is clear that all the operators 
which contains $A$ annihilate $|S\rrangle$.
Therefore the generators which give rise to new states
are limited to $J_n$ ($n=1,2,3,\cdots$).  The general singlet state
takes the form $P(J_1,J_2,\cdots)|S\rrangle$
where $P$ is an arbitrary polynomial of $J_1,J_2,\cdots$.
This is the Fock space of a free boson field or equivalently a
free fermion field.
It is consistent with the
form of the partition function\footnote{The level $n$ states are written as 
$J_{n_1}J_{n_2}\cdots |S\rrangle \;\;(n_1+n_2+\cdots=n;~n_1\geq n_2 \geq
\cdots \geq 0)$. The number of these states is given by the number of partitions of $n$, and $Z^{(S)}(q)$ is well-known 
as the generating function of this partition number.}
$Z^{(S)}(q)$ in (\ref{part1}).
For the commuting charges $H_n$, the weights of
$|S\rrangle$ are
\begin{eqnarray}
{H_m} |S\rrangle=0\;\;\;\; (m=1,2,3,\cdots).
\end{eqnarray}

\paragraph{Highest weight state at level 1 and the adjoint sector}
At this level, there are $N^2$ states $\bar A_{ij}|0\rangle$.
The only nontrivial highest weight condition
is $\mbox{Tr}(A)|\mbox{state}\rangle=\cO^{(-)}|\mbox{state}\rangle=0$.
The solutions to this condition are $N^2-1$ states,
\begin{eqnarray}\label{adjoint1}
 |A_1\rrangle_{ij}=(\bar{A}_{ij}-\frac{1}{N}\delta_{ij}
\mbox{Tr}\bar A)|0\rangle\,,\quad
(i,j=1,\cdots,N)\,.
\end{eqnarray}
It is easy to see that $|A_1\rrangle$ transforms as the
adjoint representation,
\begin{equation}
J_\Lambda|A_1\rrangle_{ij}=
\sum_k\left(\Lambda_{ik}|A_1\rrangle_{kj}
-\Lambda_{kj}|A_1\rrangle_{ik}\right)\,.
\end{equation}
The remaining one state at level $1$ is a singlet state
generated from $|S\rrangle$.
As we expected, the highest weight condition
of $\hW_\infty$ automatically picks up an
irreducible representation of $U(N)$.

It is easy to see the operators which contain more than one
$A$ annihilate $|A_1\rrangle$.  On the other hand applying
$L_n$ generates a new state,
\begin{eqnarray}
 L_n\left(\bar{A}_{ij}
-\frac{1}{N}\delta_{ij}\mbox{Tr}(\bar A)\right)|0\rangle=
\left((\bar{A}^{n+1})_{ij}
-\frac{1}{N}\delta_{ij}\mbox{Tr}(\bar A^{n+1})\right)|0\rangle\,.
\end{eqnarray}
Applying $L_{n'}$  further produces the state of the same form
with different $n$. Then we apply $J_n$ to generate new states.
The general states which belongs to the adjoint
representation thus takes the form,
\begin{equation}
\left((\bar{A}^n)_{ij}
-\frac{1}{N}\delta_{ij}\mbox{Tr}(\bar A^n)\right)
P(J_1,J_2,\cdots)|0\rangle
\label{eq:adj_ex}
\end{equation}
with $n=1,2,\cdots$. It coincides with the claim of \cite{r:OM}.
The number of states generated by \eqref{eq:adj_ex} is given by the generating function
\begin{align}
	\left(\sum_{n=1}^\infty q^n \right) \prod_{n=1}^\infty \frac{1}{1-q^n} 
	=\frac{q}{1-q}\prod_{n=1}^\infty \frac{1}{1-q^n}\,.
\end{align} 
This is just the character in the adjoint 
sector\footnote{Here we take a limit $N \to \infty$
in the character. At finite $N$, $\bar A^N$ can be
written in terms of $\bar A,\cdots,\bar A^{N-1}$
and we recover the finite $N$ character
\eqref{part1} and \eqref{part2}.}.
We note that all states which belong to
the adjoint representation are created by the action of the $\cW_\infty$
operators with the positive degree on the highest weight state (\ref{adjoint1}).
The eigenvalues of the highest weight states for $H_n$ are,
\begin{eqnarray}
&& {H_1}|A_1\rrangle=|A_1\rrangle\,,\quad
 {H_n}|A_1\rrangle=0\,,(n>1)\,.
\end{eqnarray}

\paragraph{Highest weight state at level 2 and  $B_2$ and $C_2$ sectors}
Level 2 highest weight state  is given by
combining $\bar A_{ij}\bar A_{kl}$ with other terms
where we take contractions of indices of this expression.
Nontrivial constraints come from $\cO^{(-)}$, $\cO^{(+--)}$
and $\cO^{(--)}$.  Actually the third one does not 
produce independent constraint
since $\left[\cO^{(-)},\cO^{(+--)}\right]=\cO^{(--)}$.
We use,
\begin{eqnarray}
 \cO^{(-)}\bar A_{ij}\bar A_{kl}|0\rangle&=&\delta_{ij}\bar A_{kl}|0\rangle
+\delta_{kl}\bar A_{ij}|0\rangle\,,\\
 \cO^{(+--)}\bar A_{ij}\bar A_{kl}|0\rangle&=&\delta_{il}\bar A_{kj}|0\rangle
+\delta_{kj}\bar A_{il}|0\rangle\,.
\end{eqnarray}
After some computation, we found combinations
\begin{eqnarray}
 |B_2\rrangle_{ijkl} &=& 
\left[\bar A_{ij}\bar A_{kl}+\bar A_{il}\bar A_{kj}\right.\nonumber\\
&&-\frac{1}{N+2}(\mathrm{Tr}\bar A)\left(
\delta_{ij}\bar A_{kl}+\delta_{kl}\bar A_{ij}
+\delta_{il}\bar A_{kj}+\delta_{kj}\bar A_{il}
\right)\nonumber\\
&&-\frac{1}{N+2}\left(
\delta_{ij}\bar A^2_{kl}+\delta_{kl}\bar A^2_{ij}
+\delta_{il}\bar A^2_{kj}+\delta_{kj}\bar A^2_{il}
\right)\nonumber\\
&&\left.+\frac{\delta_{ij}\delta_{kl}
+\delta_{il}\delta_{kj}}{(N+1)(N+2)}\left((\mathrm{Tr}(\bar A))^2
+(\mathrm{Tr}(\bar A^2))\right)\right]|0\rangle,\\
 |C_2\rrangle_{ijkl} &=& 
\left[\bar A_{ij}\bar A_{kl}-\bar A_{il}\bar A_{kj}\right.\nonumber\\
&&-\frac{1}{N+2}(\mathrm{Tr}\bar A)\left(
\delta_{ij}\bar A_{kl}+\delta_{kl}\bar A_{ij}
-\delta_{il}\bar A_{kj}-\delta_{kj}\bar A_{il}
\right)\nonumber\\
&&+\frac{1}{N+2}\left(
\delta_{ij}\bar A^2_{kl}+\delta_{kl}\bar A^2_{ij}
-\delta_{il}\bar A^2_{kj}-\delta_{kj}\bar A^2_{il}
\right)\nonumber\\
&&\left.+\frac{\delta_{ij}\delta_{kl}
-\delta_{il}\delta_{kj}}{(N+1)(N+2)}\left((\mathrm{Tr}(\bar A))^2
-(\mathrm{Tr}(\bar A^2))\right)\right]|0\rangle\,,
\end{eqnarray}
satisfy the highest weight condition.
The eigenvalues of $H_n$ for these states are,
\ba
&& H_1|B_2,C_2\rrangle=2|B_2,C_2\rrangle\,,\quad
 H_2|B_2\rrangle=2|B_2\rrangle\,,\quad
 H_2|C_2\rrangle=-2|C_2\rrangle\,,\nn\\
 &&H_m|B_2,C_2\rrangle=0\,,\quad(m>2)\,.
\ea
We note that $H_2$ eigenvalue distinguishes $|B_2\rrangle$
and $|C_2\rrangle$ which could not be separated
by $\cO^{(-)}$ and $\cO^{(+--)}$ conditions alone.

For these states the application of $V_{n,m}$ becomes
nontrivial. It produces the states of the form,
\begin{eqnarray}
\left( (\bar A^n)_{ij}(\bar A^m)_{kl}+
 (\bar A^m)_{ij}(\bar A^n)_{kl}\pm
  (\bar A^n)_{il}(\bar A^m)_{kj}\pm
 (\bar A^n)_{kj}(\bar A^m)_{il}+\cdots\right)|0\rangle
 \label{b2c2g}
\end{eqnarray}
for $B_2$ (resp. $C_2$) representations. 
It is easy to check that applying $V_{n',m'}$ further, 
or $L_{n'}$ does not produce new type of  states. 
Since they are
symmetric under $n\leftrightarrow m$, the number of the state
which is generated from the first factor is given as,
\begin{equation}
 \sum_{1\leq n\leq m}q^{n+m}=\frac{q^2}{(1-q)(1-q^2)}\,.
\end{equation}
This is exactly the prefactor $P^{(B_2,C_2)}(q)$ in (\ref{part2})
in the large $N$ limit.
Finally we can multiply the state with any polynomials of $J_n$
as the singlet and the adjoint sectors. So the most general form of
$B_2$ and $C_2$ representation sector has the form (\ref{b2c2g})
multiplied by $P(J_1,J_2,\cdots)$.

We note that at level two, there are extra states,
$2$ singlet states ($J_1^2|S\rrangle$
and $J_2|S\rrangle$),
$2(N^2-1)$ states with adjoint representation
($J_1|A_1\rrangle$ and $L_1 |A_1\rrangle$), and $d_{B_2}, d_{C_2}$ states with $B_2$ and $C_2$ representation respectively.
These span precisely the $N^2(N^2+1)/2$ dimensional
level two states.

We have seen that the highest weight conditions of the $\hW_\infty$ algebra
are useful to obtain the explicit form of the wave functions in the non-singlet sectors.
It is also clear that the number of the state at each level is given by the
character of the $\hW_\infty$ algebra.  We do not have, however, the complete classification
of the irreducible representations of $\hW_\infty$ which should be classified
according to the eigenvalues of $H_n$.  This issue remains as an important open problem
which should be solved as the $\cW_\infty$ algebra \cite{r:Kac,r:AFMO}.

\section{Calogero(-Sutherland)-type interaction in  $\hW_\infty$ generators}
\label{sec:eigen}
In the following chapters, we derive the exact spectrum of
the commuting charges $H_n$ of the $\hW_\infty$ algebra
and their eigenstates.  Usually, this problem is approached
by reducing the dynamical degree of freedom to the eigenvalues
of matrices $X$ and solve the Hamiltonian problem
with the Calogero-type interaction (\ref{cano}).
As we noted in the previous section, 
in the creation and annihilation basis, to find
eigenfunction of the Hamiltonian is trivial since
the Hamiltonian just counts the number of creation operators
applied to the vacuum.  Therefore a nontrivial issue is
how to diagonalize higher charges $H_n$ ($n=2,3,4,\cdots$).
In this section, we derive the explicit form
of $H_2$ in terms of the eigenvalues of $\bar A$
and derive the interaction term which is similar
to the Calogero-Sutherland interaction.
We will solve the eigenvalue problem (for the adjoint sector)
in the next section by applying the bosonization technique.
The procedure in these sections  illuminates
the direct correspondence between the solvable system and
CFT which generalizes the free fermion in the singlet sector. 

Since $A$ and $\bar A$ are $q$-numbers, it will be more convenient
to introduce $c$-number matrix $\bar Z$ as the eigenvalue
of $\bar A$ in the coherent state basis,
\begin{equation}
 \langle 0|e^{\mathrm{Tr}(A \bar Z)}|\Psi\rangle=\Psi(\bar Z)\,\,,
\quad \bar A_{ij}|\Psi\rangle\rightarrow \bar Z_{ij}\Psi(\bar Z)\,,\quad
A_{ij}|\Psi\rangle \rightarrow 
\frac{\partial}{\partial \bar{Z}_{ji}}\Psi(\bar Z)\,.
\end{equation}

We diagonalize $\bar Z$ as $\bar Z=e^K \bar z e^{-K}$
where $\bar z$ is a diagonal matrix and $e^K$ ($K:$ anti-hermitian) is a unitary
transformation which is needed for the diagonalization. 
We expand  $\bar Z$ in terms of $K$,
\begin{equation}
 \bar Z_{ij}=\bar z_{i}\delta_{ij}+(\bar z_j-\bar z_i)K_{ij}+\frac{1}{2}
\sum_k(\bar z_i+\bar z_j -2\bar z_k)K_{ik}K_{kj}+\mathcal{O}(K^3)\,.
\end{equation}
$K$ will be put to be zero at the end of the computation.
However, in order to express the differentiation with respect to the
matrix $\bar Z$, we need to keep it for a while.

We determine the differentiation with respect to $\bar A$
by the requirement,
\begin{equation}\label{matdiff}
 \frac{\partial}{\partial \bar Z_{ij}}
\bar Z_{kl}=\delta_{ik}\delta_{jl}\,.
\end{equation}
We write 
\begin{equation}
 \frac{\partial}{\partial \bar Z_{ij}}=
\sum_{r}\mathcal{A}_{ij,r}\partial_r
+\sum_{r,s}\mathcal{B}_{ij,rs}\frac{\partial}{\partial K_{sr}}
\end{equation}
(with $\partial_r=\frac{\partial}{\partial \bar z_r}$)
and expand the coefficients as 
\begin{equation}
 \mathcal{A}_{ij,r}= \mathcal{A}^{(0)}_{ij,r}+ \mathcal{A}_{ij,r}^{(1)}
+ \mathcal{A}_{ij,r}^{(2)}\cdots\,,\quad
 \mathcal{B}_{ij,r}= \mathcal{B}^{(0)}_{ij,rs}+ \mathcal{B}_{ij,rs}^{(1)}
+ \mathcal{B}_{ij,rs}^{(2)}\cdots\,,
\end{equation}
where $\mathcal{A}^{(n)}$ and $\mathcal{B}^{(n)}$ are
$\mathcal{O}(K^n)$.
We can fix these coefficients order by order 
by requiring (\ref{matdiff}). For example,
the coefficients for $\cO(1)$ and $\cO(K)$ are,
\begin{eqnarray}
 && \mathcal{A}^{(0)}_{ij,k}=\delta_{ij}\delta_{jk}\,,\\
 && \mathcal{B}^{(0)}_{ij,rs}=
\frac{\delta_{ir}\delta_{js}(1-\delta_{rs})}{\bar z_r-\bar z_s}\,,\\
&& \mathcal{A}^{(1)}_{ij,k}=K_{ij}(\delta_{jk}-\delta_{ik})\,,\\
&& \mathcal{B}^{(1)}_{ij,lk}=\left(
\frac{1-\delta_{kl}}{\bar z_l-\bar z_k}-\frac{1}{2}
\frac{1-\delta_{ik}}{\bar z_i-\bar z_k}
\right)\delta_{jk}K_{il}-
\left(
\frac{1-\delta_{kl}}{\bar z_l-\bar z_k}-\frac{1}{2}
\frac{1-\delta_{ij}}{\bar z_i-\bar z_j}
\right)\delta_{il}K_{kj}
\end{eqnarray}
It is equivalent to the expression,
\begin{eqnarray}
 \frac{\partial}{\partial \bar Z_{ij}}&=&\delta_{ij}\partial_j
+\frac{1-\delta_{ij}}{\bar z_i-\bar z_j}
\partial_{K_{ji}}+ K_{ij}(\partial_j-\partial_i)
+\sum_l\left(\frac{1-\delta_{jl}}{\bar z_l-\bar z_j}-\frac{1}{2}
\frac{1-\delta_{ij}}{\bar z_i-\bar z_j}\right)K_{il}\partial_{K_{lj}}
\nonumber\\
&&-\sum_l\left(\frac{1-\delta_{il}}{\bar z_i-\bar z_l}-\frac{1}{2}
\frac{1-\delta_{ij}}{\bar z_i-\bar z_j}\right)K_{lj}\partial_{K_{li}}
+\mathcal{O}(K^2)\,.
\end{eqnarray}
These operators can be applied to the wave function with
the representation $\rho$ through the $K$ dependence in
$\Psi^{(\rho)}_a(\bar Z)=\rho(e^K)_{ab}\Psi^{(\rho)}_b(\bar z)$.
In order to see it more explicitly, we consider the adjoint representation
in the following. The wave function  becomes,
\begin{eqnarray}
 \Psi(\bar Z)_{ij}&=&\left(e^K \Psi(\bar z)e^{-K}\right)_{ij}\nonumber\\
& =&\psi_i\delta_{ij}+(\psi_{j}-\psi_i)K_{ij}
+ \frac{1}{2}\sum_k(\psi_i+\psi_j-2\psi_k)K_{ik}K_{kj}+\cdots
\end{eqnarray}
where we denote the diagonal component of $\Psi(\bar z)$ 
as $\psi_i(\bar z)$.
The expression for the higher charges in terms of the eigenvalues
are given by writing the action of 
$\mbox{Tr }(\bar Z\frac\partial{\partial \bar Z})^n$
to $\Psi(\bar Z)$ and putting $K=0$ at the end.
We note that in the course of the computation we need higher
$K$ dependent terms since we have differentiation with respect to $K$.

Since we have an explicit expression only to the first order in
$K$, we can obtain only ${H}_1$ and $H_2$.
The first one $H_1$ is trivial,
\begin{equation}
 (H^{(A_1)}_1\cdot\psi)_k=\sum_i {\bar z_i\partial_i}\psi_k\,.
\end{equation}
We put a suffix $A_1$ in order to specify that this is the expression for the
adjoint sector.
The second one becomes,
\begin{eqnarray}
&&(H^{(A_1)}_2\cdot\psi)_k=\sum_i(\bar z_i\partial_i)^2 \psi_k
-2\sum_{i(\neq k)}\frac{\bar z_i\bar z_k}{(\bar z_k-\bar z_i)^2}
(\psi_k-\psi_i)\nonumber\\
&&~~~-\sum_{i}(\bar z_i\partial_i) \psi_k
+\sum_{i\neq j}\left(
\bar z_i\bar z_j\frac{1-\delta_{ij}}{\bar z_i-\bar z_j}
(\partial_i-\partial_j)
\right)\psi_k
\label{h2}
\end{eqnarray}
The first two terms give the Calogero-Sutherland Hamiltonian.
The third term is proportional to $H_1$ and not relevant for
the diagonalization.  The fourth term is somewhat new.
If we rewrite  $\bar z_i=e^{\theta_i}$,
 the action of $H_2$ can be written as Calogero-Sutherland \cite{r:CS}
like form,
\begin{eqnarray}
 (H^{(A_1)}_2\psi)_i&=&\sum_i\frac{\partial^2 \psi_k}{\partial\theta_i^2}
-\sum_i\frac{\partial\psi_k}{\partial \theta_i}
-\frac{1}{2}\sum_{i(\neq k)}
\frac{\psi_k-\psi_i}{\mathrm{sinh}^2(\theta_{ik}/2)}
\nonumber\\
&&+\frac{1}{2}\sum_{i\neq j}\frac{1}{\sinh (\theta_{ij}/2)}
\left(e^{-\theta_{ij}/2}\partial_{\theta_i}
-e^{\theta_{ij}/2}\partial_{\theta_j}\right)\psi_k,
\label{h2s}
\end{eqnarray}
where $\theta_{ij}=\theta_i-\theta_j$.

For the usual Calogero-Sutherland model, there have been
a large number of references where Hamiltonian
system analogous to (\ref{h2}) or (\ref{h2s}) is solved 
directly in terms of $\bar z$ variables \cite{r:Poly}.
Here, rather than moving to this direction,  
we rewrite this Hamiltonian by bosonization technique and 
solve it from this approach \cite{r:AMOS}.

\section{Bosonization}
\label{sec:boson}
In order to describe the non-singlet sectors  in a way similar to
the singlet sector, it is natural to introduce free boson variables,
\begin{equation}
 p_r=\mbox{Tr }\bar Z^r=\sum_{i}\bar z^r_i\,,
\end{equation}
(which is
called ``collective coordinate''  or ``power sum''
in the literature)
and rewrite the $\hW_\infty$ operators in terms of $p_r$
and the degree of freedom associated with the tips.
Free boson oscillators are usually  defined as
$\alpha_n=n\partial_{p_n}$, $\alpha^\dagger_{n}=p_n$
for $n>0$.  They satisfy  standard commutation relations,
\begin{equation}
 \left[\alpha_n,\alpha^\dagger_m\right]=n\delta_{n,m}\,.
\end{equation}

In the following we focus on writing the explicit forms of 
the quantities $H_1$, $H_2$, $H_3$ when they are applied to 
specific (non-)singlet sectors. For simplicity
we consider the singlet $(S)$, the adjoint $(A_1)$,
and $B_2$, $C_2$ representations.
For each case, by the results of \S\ref{sec:Winf},
the wave functions are written as the linear combinations
of 
\begin{eqnarray}
{S}&:& f(p_1,p_2,\cdots)\quad \Rightarrow \quad f(p)\nonumber\,,\\
{A_1}&:& ((\bar Z^n)_{ij}+\cdots) f(p_1,p_2,\cdots)
\quad \Rightarrow \quad f(p)|n\rangle\,,\nn\\
{B_2}&:&\left( (\bar Z^n)_{ij}(\bar Z^m)_{kl}+
 (\bar Z^m)_{ij}(\bar Z^n)_{kl}+
  (\bar Z^n)_{il}(\bar Z^m)_{kj}+
 (\bar Z^n)_{kj}(\bar Z^m)_{il}+\cdots\right)f(p_1,p_2,\cdots)\nn\\
&& \quad \Rightarrow \quad f(p)|n,m;+\rangle\,,\nn\\
{C_2}&:&\left( (\bar Z^n)_{ij}(\bar Z^m)_{kl}+
 (\bar Z^m)_{ij}(\bar Z^n)_{kl}-
  (\bar Z^n)_{il}(\bar Z^m)_{kj}-
 (\bar Z^n)_{kj}(\bar Z^m)_{il}+\cdots\right)f(p_1,p_2,\cdots)\nn\\
&& \quad \Rightarrow \quad f(p)|n,m;-\rangle\,.
\label{states}
\end{eqnarray}
Here we have skipped writing the subleading order terms
since they are not relevant in the following computation.
In order to keep the formulae as simple as possible, 
we will use short
hand notations which are written after the arrow $\Rightarrow$.
The states $|n\rangle$, $|n,m;\pm\rangle$ represent the 
degrees of the freedom associated with the tips.
For $B_2$ and $C_2$, they are 
symmetric $|n,m;\pm\rangle=|m,n;\pm\rangle$.

There are two different paths to obtain a
bosonic representation of  $H_n$. One is to use the expressions
in terms of the eigenvalues which are obtained
in the previous section.  This approach has a benefit
in showing the explicit relation between  the Calogero-Sutherland
type interactions and  their bosonized representations.
Another approach is to work directly with the matrix variables.  
Both approaches give, of course, the same answer.
Since the expressions in terms of the eigenvalues
are getting more and more complicated for the
higher charges and the higher representations, 
we will basically use the latter approach.

The computation itself is straightfoward.  We apply
$\bar Z$ representation of $H_n$
\begin{eqnarray}
H_1&=&\sum_{i,j} \bar Z_{ij} \frac{\partial}{\partial \bar Z_{ij}}\,,
\quad H_2=\sum_{i,j,k,l} \bar Z_{ij} \bar Z_{kl}
\frac{\partial}{\partial \bar Z_{kj}}
\frac{\partial}{\partial \bar Z_{il}}\,,\nonumber\\
H_3&=&\sum_{i,j,k,l,n,m} \bar Z_{ij} \bar Z_{kl}\bar Z_{nm}
\frac{\partial}{\partial \bar Z_{kj}}
\frac{\partial}{\partial \bar Z_{nl}}
\frac{\partial}{\partial \bar Z_{im}}\,,
\end{eqnarray}
to the states (\ref{states}) and rewrite the
results by using the multiplications  and the derivations 
of $p_r$.  The useful formulae are, for example,
\begin{eqnarray}
&&	\left[\frac{\partial}{\partial \bar
	Z_{ji}},(\bZ^n)_{kl}\right]=\sum_{m=1}^{n} 
(\bZ^{m-1})_{kj} (\bZ^{n-m})_{il}\,,\nn\\
&&	\left[\frac{\partial}{\partial \bar
	Z_{ji}},(p_k)^n\right] =n k (\bZ^{k-1})_{ij}(p_k)^{n-1}, \quad
	\left[\frac{\partial}{\partial \bar
	Z_{ji}}, f(p)\right]=\sum_{r=1}^N
 r (\bZ^{r-1})_{ij} \pd_{p_r}f(p)\,.
 \end{eqnarray}
After some computation, we arrive at the bosonized formulae
for $H_{1}$,$H_{2}$, $H_{3}$.

The expressions for $H_1$ (Hamiltonian) are trivial as usual,
\begin{eqnarray}
 H^{(S)}_1 f(p) & = & \left(
\sum_r rp_r\partial_{p_r}\right)f(p) \,,\\
 H^{(A_1)}_1 f(p)|n\rangle & = & \left(
n+\sum_r rp_r\partial_{p_r}\right)f(p)|n\rangle \,,\label{eq:H1act}\\
 H^{(B_2,C_2)}_1 f(p)|n,m;\pm\rangle & = & \left(
n+m+\sum_r rp_r\partial_{p_r}\right)f(p)|n,m;\pm\rangle \,.
\end{eqnarray}
Here and in the following, 
we use the upper sign for $B_2$ and the lower sign for $C_2$.

In the expressions for $H_2$, there appear  various cross terms
among the free bosons and the tips, 
\begin{eqnarray}
{H^{(S)}_2} f(p) & = & 
\left(
\sum_{r,s=1}^Nrs p_{r+s}\partial_{p_r}\partial_{p_s}
+\sum_{l=1}^N\sum_{r=1}^{l-1}lp_rp_{l-r}\partial_{p_{l}}
\right)f(p) \,,\\
{H^{(A_1)}_2} f(p)|n\rangle & = & 
2n\sum_{r=1}^N r\partial_{p_r} f(p)|n+r\rangle
+2\sum_{s=1}^{n-1} (n-s)p_s f(p)|n-s\rangle\,,\nn\\
&&+ H_2^{(S)}f(p) |n\rangle\,,
\label{eq:H2act}\\
{H^{(B_2,C_2)}_2} f(p)|n,m;\pm\rangle & = & 
2n\sum_{r=1}^N r\partial_{p_r} f(p)|n+r,m;\pm\rangle
+2m\sum_{r=1}^N r\partial_{p_r} f(p)|n,m+r;\pm\rangle\nn\\
&&+2\sum_{s=1}^{n-1} (n-s)p_s f(p)|n-s,m;\pm\rangle
+2\sum_{s=1}^{m-1} (m-s)p_s f(p)|n,m-s;\pm\rangle
\nonumber\\
&&
\pm 2 \sum_{r=1}^n\sum_{s=1}^m f(p)
|n-r+s,m+r-s;\pm\rangle
+ H_2^{(S)}f(p)|n,m;\pm\rangle\,.
\end{eqnarray}
Obviously $H_2^{(S)}$ describes the mixing among
the free bosons.  In the free fermion language, it reduces to 
$\oint\bar\psi(z)(z\partial_z)^2\psi(z)$.
This part is common in every sector.
The first two terms in $H_2^{(A_1)}$
and the first four terms in $H_2^{(B_2,C_2)}$
describe the mixing between the tip and the free boson.
They have again the similar form in both $A_1$ and $B_{2}, C_{2}$.
Finally the fifth term in $H_2^{(B_2,C_2)}$ describes the
mixing among the tips. We note the sign difference
between $B_2$ and $C_2$.
It is not difficult to confirm that these interactions take
the similar forms for even higher representations.

These expressions can be written without the reference
to the wave functions
if we  introduce the shift operators $\hat E_p$ as
(for the adjoint sector)
\begin{equation}
 \hat E_{p}|n\rangle=\left\{
\begin{array}{c l}
(n-p)|n-p\rangle
 & p<n\\
0 & p\geq n
\end{array}
\right.
\,,\quad
\hat E_0|n\rangle = n | n\rangle\,,\quad
 \hat E_{-p}|n\rangle=n|n+p\rangle\,.
\end{equation}
Together with the free boson oscillator,
the conserved charges are written as
\begin{eqnarray}
 H^{(A_1)}_1   =  
 \hat E_0+H_1^{(S)}\,,\quad
{H^{(A_1)}_2}  = 
2\sum_{r=1}^N \hat E_{-r}\alpha_{r}
+2\sum_{s=1}^{N}\hat E_{s}\alpha^\dagger_s
 + H_2^{(S)}
\,,
\end{eqnarray}
where
\ba
&&H^{(S)}_1=\sum_{r=1}^N \alpha^\dagger_r\alpha_r\,,\quad
H_2^{(S)}=\sum_{r,s=1}^N\alpha^\dagger_{r+s}\alpha_{r}\alpha_s
+\sum_{l=1}^N\sum_{r=1}^{l-1}\alpha^\dagger_r\alpha^\dagger_{l-r}\alpha_l\,.
\ea
Writing a similar formula for $B_2$ and $C_2$ is straightfoward.

The expression for $H_3$ becomes more complicated
and we present only the expressions for the singlet and the
adjoint sectors.
The strategy of the computation is the same as before. 
The final result is, for the singlet sector,
\begin{equation}
	\begin{split}
	&{H_3}^{(S)}f(p)=
	\sum_{r,s,t}rst p_{r+s+t}\pd_{p_r}\pd_{p_s}\pd_{p_t}f
	-\sum_{r,s}rs p_r p_s \pd_{p_r} \pd_{p_s} f \\
	&~~~~~+\sum_{r,s}\left( \sum_{l=1}^{r+s-1}+\sum_{l=1}^{r-1} \right) rs p_l p_{r+s-l}\pd_{p_r}\pd_{p_s}f
	+\sum_r \sum_{l=1}^{r-1}\sum_{m=1}^{r-l-1} r p_l p_m p_{r-l-m} \pd_{p_r} f \\
	&~~~~~+\frac{1}{2}\sum_r r(r-1)(r-2) p_r \pd_{p_r} f 
	+\frac{1}{2} n(n-1)(n-2)f \,.
	\end{split}
\end{equation}
The expression for the adjoint sector involves extra terms
which describe the mixing between the bosons and the tip,
\begin{equation}
	\begin{split}
	&{H_3}^{(A_1)}f(p)\ket{n}=(H_3^{(S)}f(p))\ket{n}\\
	&~~~~~+3n \sum_{r,s}rs \pd_{p_r}\pd_{p_s}f \ket{n+r+s}
	+3n\sum_r \sum_{l=1}^{r-1}r p_l\pd_{p_r} f \ket{n+r-l} \\
	&~~~~~+3\sum_r \sum_{l=1}^{r-1}r (n-l)p_l \pd_{p_r} f \ket{n+r-l} 
	+3\sum_r \sum_{l=1}^{n-1} r(n-l)p_{r+l}\pd_{p_r} f \ket{n-l} \\
	&~~~~~+\sum_{l=1}^{n-1}\sum_{m=1}^{l-1} (n-l)p_m p_{l-m} f \ket{n-l}
	+\sum_{l=1}^n \sum_{m=1}^{l-1} \left( \sum_{k=1}^{l-m-1} 
	+\sum_{k=1}^{n-l} \right) p_m p_k f \ket{n-m-k}\,.
	\end{split}
	\label{eq:H3_act}
\end{equation}

\paragraph{Numerical evaluation of the spectrum of $H_2$ and $H_3$}
One of the important questions in this paper is to determine 
the spectrum of $H_n$ ($n=1,2,3,\cdots$) for the various sectors.
For the singlet sector, the problem is already solved
since it is the free fermion system.
The eigenstate is labelled by Young diagram $Y$ which represents
the spectrum of fermionic system through Maya diagram 
correspondence \cite{r:fermion}.
We write the fermionic state that corresponds to $Y$ as $|Y\rangle$.
The eigenstate in terms of $p$
is written by using the boson-fermion correspondence
\be
\langle 0|e^{\sum_{n=1}^\infty \frac{\alpha_n}{n}p_n}|Y\rangle=s_Y(p)
\ee
where $s_Y(p)$ is Schur polynomial in terms of the power sum.
The eigenvalues of $H_{1,2}$ is given as,
\begin{eqnarray}
 && h_1(Y) =\sum_i \mu_i=|Y|\,,\qquad
 h_2(Y)=\sum_{i=1}^l \mu_i(\mu_i-2i+1)
\label{h2Y}
\end{eqnarray}
where $\mu_i$ is the number of boxes in $i$-th raw in $Y$
($\mu_1\geq \mu_2\geq\cdots\geq \mu_l>0$).
The eigenvalues of $H_2^{(S)}$ and their multiplicity are
summarized in the following table. In the table, ``level'' 
means the eigenvalue of $H_1$.
One can see that the most of the degeneracy of the spectrum
at the level of the Hamiltonian ($H_1$) is resolved by considering
$H_2$.
Some eigenvalues have the multiplicity larger than one since 
different Young tableaux have the same values of $h_2$ accidentally.
These degeneracies are resolved 
by considering $H_3$ and so on.  
\begin{center}
 \begin{tabular}{|c||l|}\hline
 level & ${H^{(S)}_2}$ eigenvalues {\em (exponent is multiplicity)}\\ \hline\hline
 0 & 0\\ \hline
 1 & 0\\ \hline
 2 & $\pm 2$\\ \hline
 3 & $\pm 6$, $0$\\ \hline
 4 & $\pm 12$, $\pm 4$, $0$ \\ \hline
 5 & $\pm 20$, $\pm 10$, $\pm 4$, $0$\\ \hline
 6 & $\pm 30$, $\pm 18$, $\pm 10$, $\pm 6^2$, $0$\\ \hline
 7 & $\pm 42$, $\pm 28$, $\pm 18$, $\pm 14$, 
$\pm 12$, $\pm 6$, $\pm 2$, $0$\\ \hline
8 & $\pm 56$, $\pm 40$, $\pm 28$, $\pm 24$,
 $\pm 20$, $\pm 16$, $\pm 14$, $\pm 8^2$, $\pm 4$, $0^2$\\
\hline
 \end{tabular}
\end{center}

For the adjoint sector (and also $B_2$ and
$C_2$ sectors), it is still difficult for us to evaluate
the spectrum in terms of the bosonization language.
We will determine it from the different viewpoint
in the next section. Here we present the result
in advance to explain the important feature of
the spectrum of the non-singlet sector.
It is, of course, possible to perform a direct computer
calcuation based on 
\eqref{eq:H1act} and \eqref{eq:H2act}
(actually this is what we did before we found the analytic solution).
These analysis are, of course, consistent with each other.

The spectrum of $H_2^{(A_1)}$ is shown by the following table.
%
\begin{center}
 \begin{tabular}{|c||l|}\hline
 level & ${H^{(A_1)}_2}$ eigenvalues {\em (exponent is multiplicity)}\\ \hline\hline
 1 & 0\\ \hline
 2 & $\pm 2$\\ \hline
 3 & $\pm 6$, $0^2$\\ \hline
 4 & $\pm 12$, $\pm 4^2$, $0$ \\ \hline
 5 & $\pm 20$, $\pm 10^2$, $\pm 4^2$, $0^2$\\ \hline
 6 & $\pm 30$, $\pm 18^2$, $\pm 10^2$, $\pm 6^3$, $0^3$\\ \hline
 7 & $\pm 42$, $\pm 28^2$, $\pm 18^2$, $\pm 14^2$, 
$\pm 12^2$, $\pm 6^3$, $\pm 2^2$, $0^2$\\ \hline
8 & $\pm 56$, $\pm 40^2$, $\pm 28^2$, $\pm 24^2$,
 $\pm 20^2$, $\pm 16$, $\pm 14^3$, $\pm 8^5$, $\pm 4^2$, $0^5$\\
\hline
 \end{tabular}
\end{center}
Interestingly, the eigenvalues are precisely 
identical with the singlet sector (\ref{h2Y}) up to the multiplicity!
It is surprising that the extra terms in $H^{(A_1)}_2$ associated 
with the ``tip'' does not change the spectrum. 

It may be a natural guess that the degeneracy appearing 
$H_2$ in the adjoint sector can be removed if 
we consider higher conserved charges
such as $H_3,H_4\ldots$.   However this is not the case.  
The following table shows the eigenvalues of $H_3^{(A_1)}$ 
and their multiplicity computed from \eqref{eq:H3_act}.
\begin{center}
 \begin{tabular}{|c||l|}\hline
 level & ${H^{(A_1)}_3}$ eigenvalues {\em (exponent is multiplicity)}\\ \hline\hline
 1 & $0$\\ \hline
 2 & $0^2$\\ \hline
 3 & $6^2$, $-3^2$\\ \hline
 4 & $24^2$, $0^4$, $-12$ \\ \hline
 5 & $60^2$, $15^4$, $0^2$, $-12^4$\\ \hline
 6 & $120^2$, $48^4$, $12^4$, $0^4$, $-15^3$, $-24^2$\\ \hline
 7 & $210^2$, $105^2$, $42^4$, $30^4$, 
$21^2$, $-6^6$, $-15^4$, $-30^4$\\ \hline
8 & $336^2$, $192^4$, $96^4$, $84^4$,
 $48^4$, $21^6$, $12^4$, $0^3$, $-24^{12}$, $-48^2$\\
\hline
 \end{tabular}
\end{center}
These eigenvalues are identical to the exact spectrum $h_3(Y)$ which
we derive later \eqref{eq:H3_eigenvalue}.
We note that the multiplicity is even larger than that of $H_2^{(A_1)}$
because the eigenvalues $h_3(Y)$  have additional
degeneracy for the states associated with different Young diagrams.
%
It resolves, however, some accidental degeneracy in $H_2$.
For  example, the Young diagrams $[4,1,1]$ and $[3,3]$ have the same
$h_2(Y)$ but different $h_3(Y)$,
\begin{align}
	h_2([4,1,1])=h_2([3,3])=6,\;\;
	h_3([4,1,1])=12,\;\; h_3([3,3])=-24.
\end{align} 
The most important  degeneracy due to the location of
the ``tip''  remain.

\section{Group theoretical construction of eigenstates of $H_n$}
\label{sec:eig}
\subsection{Analogy with matrix string theory}
In the previous section, we have seen that the spectrum of
$H_2$ (and also $H_3$) for the adjoint sector
is identical to that for the singlet sector.
This feature  actually remains the same for any non-singlet
sectors.  We will show this fact by explicit construction of
their eigenstates. It turns out that the problem can be solved by
an application of the group theory at the relatively elementary level.

We first develop a compact notation to represent
generic states in the MQM Hilbert space.
We note that a state of the form
$\bar A_{i_1 j_1}\cdots \bar A_{i_n j_n}|0\rangle$
can be written as a single trace operator acting on the Fock vacuum,
$\Tr (\cC_1\bar A \cdots \cC_n \bar A)|0\rangle$
with $(\cC_1)_{ij}=
\delta_{ij_n}\delta_{ji_1}$, $(\cC_2)_{ij}=\delta_{ij_1}\delta_{ji_2}$
and so on. It is clear that arbitrary states can be written
as the sum of such single trace form with more general
constant $\cC$.  
In order to describe more generic state, 
it is more useful to introduce the multi-trace operators,
$
 \prod_{I=1}^P \mbox{Tr }(\cC^{(I)}_1 \bar A\cdots
\cC^{(I)}_{n_I} \bar A)|0\rangle\,.
$

With this notation, the wave function in
the singlet sector can be represented
by a linear combination of such operator with $\cC=1$.
To represent the adjoint state, we put one of $\cC$'s
as a traceless tensor and other $\cC$ to be identity.
For $B_2$ and $C_2$, we keep two of $\cC$'s arbitrary
with the appropriate symmetry properties.

\begin{figure}[tbp]
	\begin{center}
		\includegraphics[width=0.8\linewidth]{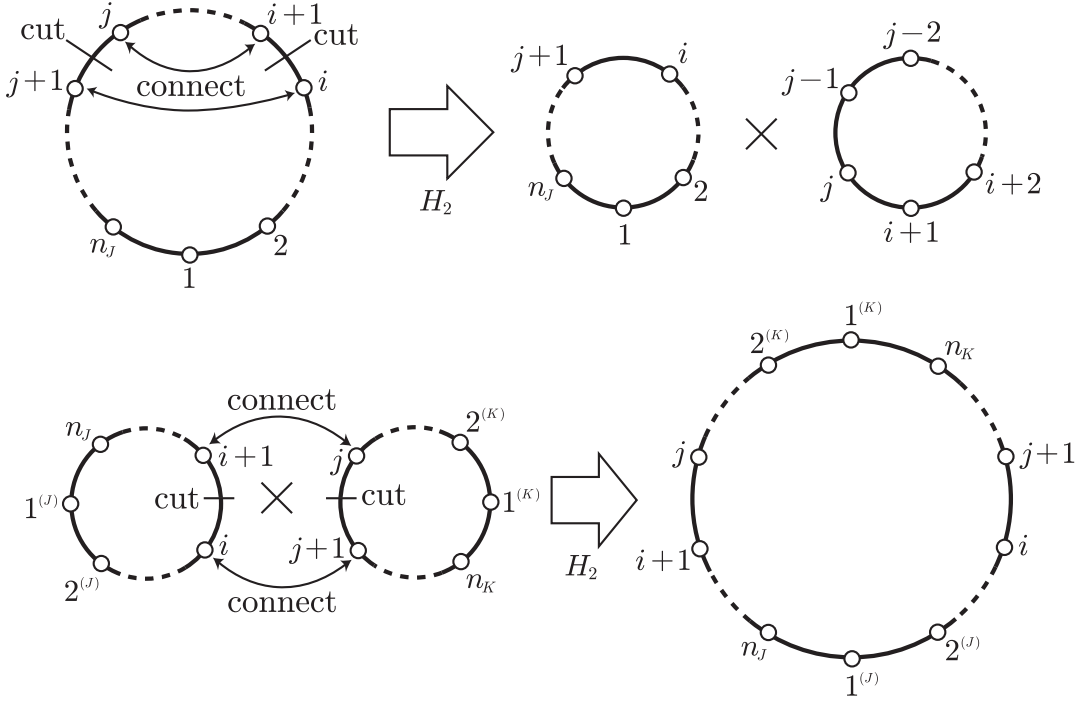}
	\end{center}
	\caption{Splitting and joining rules of $H_2$.  Each loop represents a single trace operator.}	
	\label{fig:splitting-joining}		
\end{figure}

We apply $H_2$ to such multi-trace operator state,
the an interesting structure shows up,
\begin{eqnarray}
 && \frac{1}{2}{H_2} \prod_{I=1}^P 
\mbox{Tr }(\cC^{(I)}_1 \bar A\cdots
\cC^{(I)}_{n_I} \bar A)
|0\rangle\nn\\
&&~=\sum_{J=1}^P
\left(\prod_{I(\neq J)}^P 
\mbox{Tr }(\cC^{(I)}_1 \bar A\cdots
\cC^{(I)}_{n_I} \bar A)\right)\cdot\nn\\
&&~~~\cdot 
\sum_{1\leq i< j\leq n_J}\mbox{Tr}\left(\cC^{(J)}_{i+1}\bar A\cdots 
\cC^{(J)}_j\bar A\right)\mbox{Tr }\left(\cC^{(J)}_1
\bar A\cdots \cC^{(J)}_i \bar A
\cC^{(J)}_{j+1}\bar A\cdots \cC^{(J)}_{n_J}\bar A\right)
|0\rangle\nn\\
&&~~~~+ \sum_{J<K}
\left(\prod_{I(\neq J,K)}^P 
\mbox{Tr }(\cC^{(I)}_1 \bar A\cdots
\cC^{(I)}_{n_I} \bar A)\right)\cdot\\
&&~~~~\cdot\sum_{i,j}\mbox{Tr}
\left(
\cC^{(J)}_1\bar A\cdots\cC^{(J)}_i\bar A
\cC^{(K)}_{j+1}\bar A\cdots \cC^{(K)}_{n_K}\bar A
\cC^{(K)}_{1}\bar A\cdots \cC^{(K)}_{j}\bar A
\cC^{(J)}_{i+1}\bar A\cdots\cC^{(J)}_{n_J}\bar A
\right)|0\rangle\nn\,.
\end{eqnarray}
If we regard each single trace operator as a ``loop operator'',
the action of $H_2$ represents splitting and joining
of such operators (fig.\,\ref{fig:splitting-joining}).

This is an intriguing feature of $H_2$ which have an
interpretation in terms of the string field theory.
For example in the matrix string theory, such a splitting
and joining interaction among the long strings
is triggered by the permutation
of the short strings.  
In our case, the long string is given by the matrix $\cC\bar A$
and the short string replaced by the single trace operator.
This analogy gives us a hope that the action of
$H_2$ may have similar structure as the matrix string theory\cite{r:MatStr}.
This is indeed the case.

We have a similar representation for $H_3$.
In order to simplify the notation, we introduce
\be
\cA_{ij}=
\left\{\begin{array}{ll}
\cC_{i}\bA \cC_{i+1}\bA \cdots \cC_{j}\bA\quad &\mbox{for } i<j\\
\cC_{i}\bA \cdots \cC_{n}\bA \cC_{1}\bA \cdots \cC_{j}\bA
\quad  &\mbox{for } i>j\,.
\end{array}\right.
\ee
which describe a portion of  the
trace operator $\mbox{Tr}(\cC_{1}\bA \cdots \cC_{n}\bA)$. 
The action of $H_3$ is written as
\begin{equation}
	\begin{split}
	&\frac{1}{3}H_3\prod_{I=1}^P \mbox{Tr} (\cA_{1n_I}^{(I)})\ket{0} =\\
	&\sum_{J=1}^P \left( \prod_{I(\ne J)} \mbox{Tr} (\cA_{1n_I}^{(I)} ) \right)
	\sum_{1\leq i<j<k \leq n_J} \Bigl[ \mbox{Tr} ( \cA_{i+1,j}^{(J)} \cA_{k+1,i}^{(J)}\cA_{j+1,k}^{(J)} )
	+\mbox{Tr}(\cA_{i+1,j}^{(J)})\mbox{Tr}(\cA_{j+1,k}^{(J)})\mbox{Tr}(\cA_{k+1,i}^{(J)}) \Bigr]\ket{0} \\
	+&\sum_{J<K} \left( \prod_{I(\ne J,K)} \mbox{Tr} (\cA_{1n_I}^{(I)} ) \right) 
	\sum_{\substack{1 \leq i<j \leq n_J \\ 1\leq k \leq n_K}} 
	\Bigl[ \mbox{Tr}(\cA_{i+1,j}^{(J)}\cA_{k+1,k}^{(K)}) \mbox{Tr}(\cA_{j+1,i}^{(J)})
	+\mbox{Tr}(\cA_{j+1,i}^{(J)}\cA_{k+1,k}^{(K)}) \mbox{Tr}(\cA_{i+1,j}^{(J)}) \Bigr] \ket{0} \\
	+&2\sum_{J<K<L} \left( \prod_{I(\ne J,K,L)} \mbox{Tr} (\cA_{1n_I}^{(I)} ) \right)
	\sum_{\substack{1\leq i \leq n_J \\ 1\leq j \leq n_K \\ 1\leq k \leq n_L}} 
	\mbox{Tr} (\cA_{i+1,i}^{(J)}\cA_{j+1,j}^{(K)}\cA_{k+1,k}^{(L)})\ket{0}. 
	\end{split}
\end{equation}
It describes splitting and joining processes of $3\rightarrow 1$,
$2\rightarrow 2$, $1\rightarrow 3$  between the loops.
Clearly one can continue to carry out this type of the computation
for higher $H_n$.  

\subsection{Exact eigenstates in terms of Young symmetrizer}
In the matrix string theory, the interaction between the loops
can be given by the permutation of the connection 
between the string bits.  In the following we will see that
exactly the same type of the representation exists for $H_n$
and it enables us to construct the exact eigenstate by using a
group theoretical method.

For this purpose, we introduce an economical notation of the multi-trace
operators by using the action of  the permutation group $S_n$,
\begin{equation}\label{psisigma}
 \Psi(\left\{\cC\right\};\sigma)=\sum_{a_1,\cdots,a_n}
(\cC_1\bar A)_{a_1 a_{\sigma(1)}}\cdots(\cC_n\bar A)_{a_n a_{\sigma(n)}}
|0\rangle\,.
\end{equation}
Here $\sigma$ is an element of $S_n$.
We note that a similar representation for the singlet state
apeared in \cite{r:CJR}.
The structure of the multi-trace operator comes from the
decomposition of $\sigma\in S_n$  as a product of cycles,
\begin{equation}
\sigma=(a_1a_2\cdots a_{\mu_{1}})(b_1b_2\cdots b_{\mu_2})\cdots\,,
\end{equation}
where $(a_1 a_2 \cdots a_{\mu_1})$ represents a cyclic permutation, {\it i.e.}
$\sigma(a_1)=a_2,\cdots, \sigma(a_{\mu_1})=a_1$.
The corresponding $\Psi(\left\{\cC\right\};\sigma)$ is written as
 the product of the loop operators,
\begin{equation}
 \Psi(\left\{\cC\right\};\sigma)=\mbox{Tr}(\cC_{a_1}\bar A\cdots \cC_{a_{\mu_1}}\bar A)
\mbox{Tr}(\cC_{b_1}\bar A\cdots \cC_{b_{\mu_2}}\bar A)\cdots|0\rangle.
\end{equation}
A direct computation shows that the operator $H_2$ 
can be represented as the sum of permutation operators,
\begin{equation}
 {H_2}\Psi(\left\{\cC\right\};\sigma)=\sum_{i\neq j}\Psi(\left\{\cC\right\};{\sigma(ij)})\,.
\end{equation}
As we claimed, this is the interaction of the matrix
string theory \cite{r:MatStr} where the string bits are
replaced by the matrices $\cC_i \bar A$.
More generally it is not difficult to show that the action of $H_m$  takes a 
similar  form,
\begin{equation}
 {H_m}\Psi(\left\{\cC\right\};\sigma)=\sum_{i_1,\cdots,i_m\, (i_a\neq i_b)}
 \Psi(\left\{\cC\right\};{\sigma(i_1i_2\cdots i_m)})\,,
\end{equation}
where the summation is taken for the set of mutually different integers $i_1,\cdots,i_m$.
At this point, it becomes straightfoward to construct eigenvectors
of $H_m$ by means  of the group theory.
Let us define the action of $\tau\in S_n$ on $\Psi(\sigma)$ as
the left multiplication,
\begin{equation}
 \rho(\tau)\Psi(\left\{\cC\right\};\sigma)=\Psi(\left\{\cC\right\};\tau\sigma)\,.
\end{equation}
Then it is clear that this action commute with $H_m$,
\begin{equation}
 \rho(\tau){H_m}\Psi(\left\{\cC\right\};\sigma)={H_m}\rho(\tau)\Psi(\left\{\cC\right\};\sigma)
=\sum_{i_1\cdots i_m}\Psi(\left\{\cC\right\};\tau\sigma(i_1\cdots i_m))\,.
\end{equation}
It is then possible to use Young symmetrizer  \cite{r:YounSym}
associated with
a board\footnote{
A board $B_Y$ associated with a Young diagram $Y$ is the combination of $Y$ with
the numbers, $1,2,\cdots,|Y|$ distributed in the boxes of $Y$
without overlap.  Obviously there are $|Y|!$ boards for
each Young diagram $Y$.
}  $B_Y$ of a Young diagram $Y$ to obtain an eigenstate.
The Young symmetrizer in general is written as,
\begin{equation}
 e_{B_Y}=\frac{d_Y}{n!}a_{B_Y} b_{B_Y},
\end{equation}
where
\begin{equation}
 a_{B_Y}=\sum_{\sigma \in H_{B_Y}}\rho(\sigma)\,,\quad
 b_{B_Y}=\sum_{\sigma \in R_{B_Y}}(-1)^\sigma \rho(\sigma)\,.
\end{equation}
Here $d_Y$ is the dimension of the irreducible representation of
$S_n$ associated with diagram $Y$ and $H_{B_Y}$ (resp. $R_{B_Y}$) is the
horizontal (resp. vertical) permutation associated with the board
$B_Y$.  Since $e_{B_Y}$ is a projector ($e_{B_Y}^2 =e_{B_Y}$), it defines
a projection of Hilbert space into 
 lower dimensional subspace.  We claim that the states after this projection
\begin{equation}\label{eigenstate}
 \Psi^{[B_Y]}=e_{B_Y}\Psi(\left\{\cC\right\};\sigma)\,,\quad(\sigma\in S_n)
\end{equation}
become the eigenfunctions for $H_1$, $H_2$ and $H_3$,  
\begin{equation}\label{eigenH2}
 H_1\Psi^{[B_Y]}=h_1(Y)\Psi^{[B_Y]}\,,\quad
 H_2\Psi^{[B_Y]}=h_2(Y)\Psi^{[B_Y]}\,,\quad
 H_3\Psi^{[B_Y]}=h_3(Y)\Psi^{[B_Y]}\,,
\end{equation}
where $h_1(Y)$ and $h_2(Y)$ are defined in (\ref{h2Y}) and $h_3(Y)$ will be given in
(\ref{eq:H3_eigenvalue}).
A proof of this statement is given in the next subsection
and it seems natural to conjecture that one may straightforwardly
generalize 
it for all $H_n$.

The simplicity of the form of the exact eigenstate (\ref{eigenstate}) is remarkable.
We will refer the eigenstate  constructed in this way as Young symmetrizer
state (YSS).

We note that there are some freedom in YSS
(\ref{eigenstate}), namely the choice of the constant matrices
$\cC$, the choice of the board $B_Y$ for each Young diagram $Y$
and the choice of $\sigma$.  It explains the origin of the degeneracy of
the spectrum in the non-singlet sectors.
We note that the eigenvalue depends only on the Young diagram $Y$.

Of course, the different choices  does not always 
produce independent states.  For example, 
the choice of the board can be absorbed
in the choice of the constant matrices $\cC$ assigned in each box in $Y$.  
Furthermore, the different assignments of $\cC$ sometimes
produce an identical state as we see later in the adjoint sector.  
The locations of the matrices $\cC \, (\neq 1)$ should be interpreted as
the locations of the ``tips'' in the spectrum of the background fermion.

\subsection{Calculation of the eigenvalues}
In this subsection, we give a detailed proof of (\ref{eigenH2})
for $H_2$ and $H_3$.  Proof for $H_1$ is trivial.
\paragraph{$H_2$:}
We compute the action of $H_2$ on YSS as,
\begin{equation}
 H_2\Psi^{[B_Y]}=\frac{d_Y}{n!}\sum_{i\neq j}
\Psi(a_{B_Y}  b_{B_Y} \sigma (ij))
=\frac{d_Y}{n!}\sum_{i\neq j}
\Psi(a_{B_Y}(ij)  b_{B_Y}  \sigma )\,.
\end{equation}
We can evaluate the action of transposition  $(ij)$
according to the following rule.
\begin{itemize}
 \item When the boxes $i$ and $j$ belong to the same row,
$a_{B_Y} (ij)=a_{B_Y}$.  Therefore it has eigenvalue $+1$.
There are $\sum_{i}\mu_i (\mu_i-1)$ such pairs
where $\mu_i$ is the length of $i$-th row.
 \item When the boxes $i$ and $j$ belong to the same column,
$ (ij) b_{B_Y}=-b_{B_Y}$.  Therefore it has eigenvalue $-1$.
There are $\sum_{i}\mu'_i (\mu'_i-1)$ such pairs
where $\mu'_i$ is the length of $i$-th column.
 \item When the boxes $i$ and $j$ do not meet with above two
conditions, $a_{B_Y} (ij) b_{B_Y}=0$.  It can be shown as follows.
Let $k$ be the box which belongs to the same row with
$i$ and the same column with $j$ (or vice versa),
\begin{equation}
 a_{B_Y} (ij)  b_{B_Y}= a_{B_Y} \frac{1+(ik)}{2} (ij)
 \frac{1-(jk)}{2} b_{B_Y}=0\,.
\end{equation}
\end{itemize}
The eigenvalue of $H_2$ becomes
\begin{eqnarray}
&&
\sum_i \mu_i(\mu_i-1)-\sum_i\mu'_i(\mu'_i-1)
=\left(\sum_i\mu_i (\mu_i-2i+1)\right)=\mu_2(Y)\,.
\end{eqnarray}

\paragraph{$H_3$:}
As $H_2$, we use the representation of  $H_3$ as the sum of permutation operators:
\begin{align}
	H_3 \Psi(\sigma) = \sum_{i \ne j\ne k \ne i} \Psi(\sigma  (ijk))\,.
\end{align}
The action of $H_3$ on YSS \eqref{eigenstate} becomes
\begin{align}
	H_3 \Psi^{[B]}=\frac{d_Y}{n!}\sum_{i\ne j\ne k \ne i} \Psi(a_B  b_B \sigma (ijk))
	=\frac{d_Y}{n!} \sum_{i\ne j\ne k \ne i} \Psi(a_B (ijk)  b_B \sigma)\,. 
	\label{eq:H3_action}
\end{align}
The non-vanishing contributions to the eigenvalue  come from the following three cases:
\begin{itemize}
	\item The boxes $i,j,$ and $k$ belong to the same row. In this case, since $a_B (ijk) =a_B$, the contribution
	is $\sum_i \mu_i (\mu_i-1) (\mu_i-2) \Psi^{[B]}$.
	\item The boxes $i,j,$ and $k$ belong to the same column. In this case, since $(ijk)  b_B=b_B$, the contribution
	is $\sum_i \mu_i' (\mu_i'-1) (\mu_i'-2) \Psi^{[B]}$.
 	\item Two boxes (e.g. $i$ and $j$) belong to the same row, and two boxes (e.g. $j$ and $k$) belong to the same column. 
	In this case, since $(ijk)=(ij) (jk),\, a_B (ij) =a_B$ and $(jk) b_B =-b_B$, 
	the contribution is $-3\sum_{i,j} (\mu_i-1) (\mu_j'-1) \Psi^{[B]}$.
\end{itemize}
Except for above three cases, the contribution 
of $(ijk)$ to the eigenvalue vanishes.
The reason is as follows.
Three boxes do not belong to the same row nor the same column, so we can choose two boxes that belong to 
different rows and columns.
We call these two boxes $i$ and $j$.
Let $l$ be the box which belongs to the same row with $j$ and the same column with $i$.
$l$ is not $k$ otherwise it becomes above third case.
Therefore 
\be
	a_B  (ijk) b_B = a_B\frac{1+(jl)}{2} (ji)  (jk)\frac{1-(il)}{2} b_B 
	=a_B \frac{1+(jl)}{2} (ji)\frac{1-(il)}{2}  (jk)  b_B 
	=0\,. 
\ee
Finally \eqref{eq:H3_action} becomes
\begin{align}
	H_3 \Psi^{[B]} = h_3(Y) \Psi^{[B]}\,,
	\label{eq:H3_eigenstate}
\end{align}
where
\begin{align}
	h_3(Y)=\sum_i \mu_i (\mu_i-1) (\mu_i-2) +\sum_i \mu_i' (\mu_i'-1) (\mu_i'-2)
	-3\sum_{i,j} (\mu_i-1) (\mu_j'-1).
	\label{eq:H3_eigenvalue}
\end{align}
It seems to be natural to conjecture that
generalizations of this type of the proof are possible and 
YSS is the eigenstate for higher $H_n$ ($n=4,5,\cdots$).

\subsection{Explicit forms of the exact eigenstates in terms of free boson}
The expression of the eigenfunctions (\ref{eigenstate}) as YSS  is simple,
explicit and exact.  However, in order to compare it with
the result of  CFT, it is not the most convenient form.
In the following, we present our partial result
to express YSS in the singlet and the adjoint sector
in terms of the free boson (or fermion) oscillator
and the degree of freedom associated with the tip.

In order to extract 
the eigenstate for a specific representation,
we restrict the matrices $\cC$ to a specific form.
For example in order to obtain the
singlet state, we put all $\cC$ to be identity.
To obtain the adjoint state, we put one of
$\cC$  to be a generic traceless matrix and
all the other  to be identity.
In the appendix \ref{a:lower},
we present the explicit form of the wave function
by the Young symmetrizer (\ref{eigenstate})
for the lower levels. 
It will be useful to understand the basic feature of YSS
and the discussion in this subsection.

\paragraph{The singlet sector}
For the singlet sector, it is not difficult to
prove that YSS
is identical to the Schur polynomial.  To see that,
we first observe that the wave function $\Psi^{[B_Y]}$
is unique for each Young diagram since we need to
put  all $\cC$ to be $1$.  Since $\Psi^{[B_Y]}$ and the Schur polynomial
$s_Y(J_n)|0\rangle$ are both the eigenstates of the
conserved charges $H_1,H_2,\cdots$ with the same eigenvalue
and the multiplicity , it implies
$\Psi^{[B_Y]}\propto s_Y(J_n)|0\rangle$.
The normalization is fixed by comparing the coefficient
of one particular term, for example, $J_1^n|0\rangle$.
In $\Psi^{[B_Y]}$, it comes from the term which is proportional
to $e$ in the symmetrizer and is identical  to $d_Y/n!$.
On the other hand in the Schur polynomial, it is known to be
$d_Y/n!$ again (from the Murnaghan-Nakayama rule)
\cite{r:Stan}.  We conclude,
\begin{equation}
 \Psi^{[B_Y]}|_{\cC=1}= s_Y(J_n)|0\rangle\,. \label{eq:sing-Schur}
\end{equation}
This is an  interesting relation between
the Schur polynomial and the Young symmetrizer.
One can understand this fact by reminding that 
the Young symmetrizer is the projection
to the irreducible representation when it acts on the tensor product
of the fundamental representations \cite{r:FH}.
In our case, since the state $\Psi^{[B_Y]}|_{\cC=1}$ is
given as the trace of the representation space 
associated with $Y$, it equals to the Schur polynomial, which is the character of $U(N)$.

\begin{figure}[tbp]
	\begin{center}
		\includegraphics[width=0.5\linewidth]{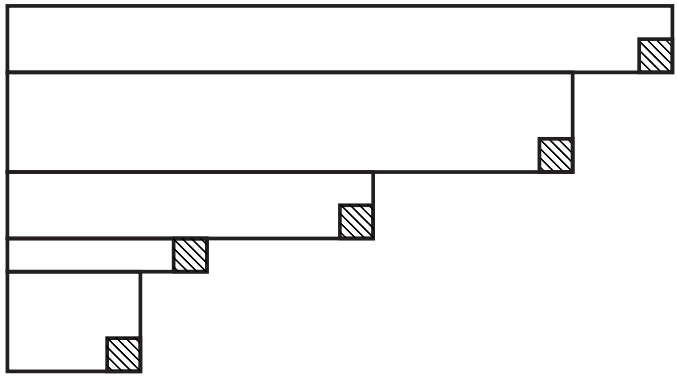}
	\end{center}
	\caption{The multiplicity for each Young diagram in the adjoint sector equals to the number of boxes 
	at the right-bottom corner of each rectangle.}	
	\label{fig:young1}		
\end{figure}

\paragraph{The adjoint sector}
In this case, the wave function $\Psi^{[B_Y]}$ depends
only the location of the box in the Young diagram $Y$
where the associated matrix $\cC$ is not $1$.
There are naively $|Y|$ choices of such boxes.
However, the number of the independent states are fewer than that. 
The multiplicity of eigenstates of $H_2$
for each Young diagram in the adjoint sector  can 
reproduced by the following rule.
We represent the Young diagram $Y$ in
the form
$[\mu_1^{r_1},\mu_2^{r_2},\cdots,\mu_s^{r_s}]$
($\mu_1>\mu_2>\cdots>\mu_s>0$).
Namely $Y$ is constructed by piling
$s$ rectangles with size $\mu_i\times r_i$ vertically.
The multiplicity for this Young diagram is
the same as the number of the rectangles, namely $s$.
It is identical to the number of the special boxes
in $Y$ where $Y$ becomes another Young diagram $Y'$ after one
removes one of these boxes.
Such boxes are located at the right-bottom corner
of each rectangle (fig.\,\ref{fig:young1}).

This rule is also consitent with the partition function.
For  the singlet states, it is written
as (one state for each Young diagram)
\begin{equation}
 \sum_{r_1,r_2,\cdots =0}^\infty q^{\sum_{s=1} ^\infty s r_s
}=\prod_{n=1}^\infty\frac{1}{1-q^n}=Z^{(S)}(q)\,.
\end{equation}
Here each set of integers $r_n$ represents a Young diagram
which consists of the rectangles with sizes $n\times r_n$.
In order to count the number of the adjoint state in the above
rule, we multiply the number of rectangles 
$m([r])=\sum_{n=1}^\infty (1-\delta_{r_n,0})$
in the summation over $r_1,r_2\cdots$.
It is then straightforward to show that
\begin{equation}
 \sum_{r_1,r_2,\cdots =0}^\infty m([r]) q^{\sum_{s=1} ^\infty s r_s
}=(q+q^2+q^3+\cdots)\prod_{n=1}^\infty\frac{1}{1-q^n}=
\frac{q}{1-q}\prod_{n=1}^\infty\frac{1}{1-q^n}
\end{equation}
which is exactly the partition function for the adjoint representation
$Z^{(A_1)}(q)$.
Physically our observation here implies that the tips in the non-singlet
sector do not affect the spectrum for any $H_n$ but only change
the multiplicity through the freedom in their locations in the Young diagram.  

From the explicit computation of the eigenfunction by the computer,
it seems rather reasonable to conjecture that 
 all the adjoint eigenstates can be written by the combinations of
 the (skew) Schur polynomials
 and the degree of freedom associated with the tip.
So far it seems difficult to write them in compact forms.
Therefore, we instead present the explicit forms of eigenstates
for the Young diagrams with the limited shape.

For the diagrams $Y=[n],[1^n]$ the wave functions are simply given by
\begin{align}
	\Psi^{[n]}=\frac{1}{n} \sum_{r=0}^{n-1} \mbox{Tr}(\cC\bA^{n-r})s_{[r]}(J)\ket{0}\,,
	\quad
	\Psi^{[1^n]}=\frac{1}{n} \sum_{r=0}^{n-1} (-1)^{n-r-1}\mbox{Tr}(\cC\bA^{n-r})s_{[1^r]}(J)\ket{0}
	\,.
\end{align}
This is a situation where only one fermion (or hole) is excited
and it is coupled with the tip.

\begin{figure}[tbp]
	\begin{center}
		\includegraphics[height=3.5cm]{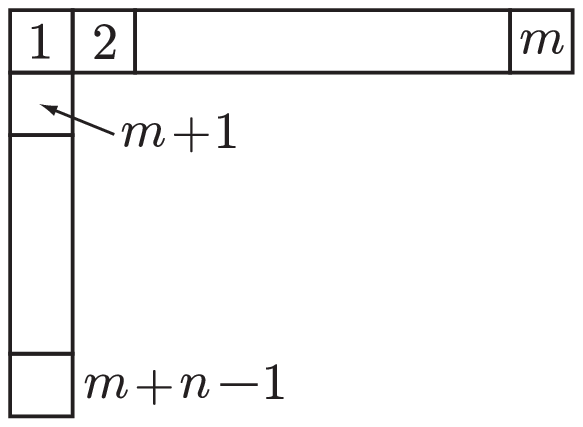}
	\end{center}
	\caption{One board with the Young tableau $[m,1^{n-1}]$.}	
	\label{fig:board}		
\end{figure}

Next we give the wave functions corresponding to $Y=[m,1^{n-1}]$.
Naively, there are three independent boards but as we argued only
two of them  are independent.
Considering the board shown in fig.\,\ref{fig:board} and putting $\cC_1=\cC$, we obtain
\begin{align}
	\Psi^{[m,1^{n-1}]}_1
	&=\frac{d_{[m,1^{n-1}]}}{(m+n-1)!} \sum_{\substack{0\leq r \leq m-1\\0\leq s \leq n-1}}(-1)^s \Tr (\cC \bA^{r+s+1})
	\frac{(m-1)!}{(m-r-1)!}\frac{(n-1)!}{(n-s-1)!} \notag \\ 
	&\hspace{120pt}\cdot (m-r-1)!s_{[m-r-1]} \cdot (n-s-1)! s_{[1^{n-s-1}]} \ket{0} \notag \\
	&=\frac{1}{m+n-1}\sum_{\substack{0\leq r \leq m-1\\0\leq s \leq n-1}}(-1)^s \Tr (\cC \bA^{r+s+1})
	s_{[m-r-1]}s_{[1^{n-s-1}]} \ket{0}.
\end{align}
The coefficient of $\Tr (\cC \bA^{r+s+1})$ comes from the number to choose $r$ boxes (considered their order)
from $2,\dots,m$ and 
$s$ boxes from $m+1,\dots,m+n-1$ and from the product of the singlet
states with remaining $Y=[m-r-1]$ and $[1^{n-s-1}]$. 
Other two states are similarly given by
\begin{align}
	\Psi^{[m,1^{n-1}]}_2=\frac{1}{(m+n-1)(m-1)}&\biggl[ \sum_{0\leq r \leq m-2}(m+n-r-2)\Tr(\cC\bA^{r+1})
	s_{[m-r-1,1^{n-1}]} \notag \\
	& +\sum_{\substack{0\leq r \leq m-2\\0\leq s \leq n-1}}(-1)^s (r+1) \Tr (\cC \bA^{r+s+2})s_{[m-r-2]}s_{[1^{n-s-1}]}
	\biggr] \ket{0} \,, \label{eq:Psi_2}\\
	\Psi^{[m,1^{n-1}]}_3=\frac{1}{(m+n-1)(n-1)}&\biggl[ \sum_{0\leq s \leq n-2}(-1)^s (m+n-s-2)\Tr(\cC\bA^{s+1})
	s_{[m,1^{n-s-2}]} \notag \\
	& +\sum_{\substack{0\leq r \leq m-1\\0\leq s \leq n-2}}(-1)^{s+1} (s+1) 
	\Tr (\cC \bA^{r+s+2})s_{[m-r-1]}s_{[1^{n-s-2}]}
	\biggr] \ket{0}\,,
\end{align}
where $\Psi^{[m,1^{n-1}]}_2$ is the state corresponding to $\cC_m=\cC$ and $\Psi^{[m,1^{n-1}]}_2$ 
is that corresponding to $\cC_{m+n-1}=\cC$.
In (\ref{eq:Psi_2}) the first term comes from the case that the boxes 1 and $m$ are not in the same cyclic 
permutation, and the second term comes from the case that both are in the same one. 
The relation among these three states is as follows:
\begin{align}\label{triag}
	(m+n-2)\Psi^{[m,1^{n-1}]}_1-(m-1)\Psi^{[m,1^{n-1}]}_2-(n-1)\Psi^{[m,1^{n-1}]}_3=0\,.
\end{align}

\section{Summary and future issues}
\label{sec:summary}
In this paper, we have studied the spectrum of MQM from the 
viewpoint of its spectrum generating algebra --- 
the $\hW_\infty$ algebra.  While the
usual $\cW_\infty$ algebra is essentially described
by free fermion systems \cite{r:Kac,r:AFMO},
$\hW_\infty$ describes systems with Calogero(-Sutherland)
type interaction which comes from  the angular part
of the matrix degree of freedom.  
The action of the $\hW_\infty$ generators has an interpretation of
the splitting and the joining of  ``loop operators''  and through
this interpretation it is possible to derive the explicit for of the 
eigenvectors for arbitrary non-singlet states.
It is remarkable that the eigenfunctions
of the commuting charges  are still classified by the Young diagram
and have the same spectrum as the free fermion  system.  
The only difference is the degeneracy of the spectrum
whose origin is the arbitrariness of the location of the tips.

There are a few questions which should be answered before
we can study the issues of the non-singlet sectors  of $c=1$ gravity.
One important aspect is how to take large $N$ limit.
While the representation in terms of the free boson and the
degree of the freedom of the tip is a good framework to take
the large $N$ analysis, our exact wave functions are defined 
through the Young symmetrizer and the translation between
the two languages seems not complete in the non-singlet
sector.

Another issue is how to solve the upside-down (UD) case.
This is in a sense obtained from the upside-up (UU) case by
a sort of Wick rotation. For instance, the wave function for
the singlet sector is the Slater determinant of the free wave functions
$x^{i\epsilon-1/2}$ (for UD case) instead of $x^n$ (for UU case).
In order to consider the matrix generalization, we need to introduce
the pure imaginary powers for the matrix $X$.
We note that the basic tools of our analysis, the $\hW_\infty$ generators,
the Young symmetrizer  and the free boson variables
are defined in terms of  the integer power of $X$.
It is clear that we need extra ideas to modify our arguments
to UD case.  We hope that the basic observation of our analysis,
the structure of the spectrum and the multiplicity
will remain the same since it is a demonstration of the fact that
the spectrum remains the same when we change the locations
of the tips.

In the mathematical side,  one important question is the
study of the  representation of the $\hW_\infty$ algebra.
While we discuss some simple irreducible representations
which appear in the context of MQM, it should be far from
the complete classification of the irreducible
representations.
The fact that the action of the generators takes the form of
splitting and joining of loop operators implies that
this algebra will be also essential to understand
string theory beyond $c=1$.  

\vskip 10mm

\noindent{\bf Acknowledgement:}
This work was started as a joint project with I. Kostov.
We are deeply grateful to him for the illuminating discussions
and for  his hospitality when Y.M. was invited to Saclay.
Y.M. is supported in part by
Grant-in-Aid (\#16540232) from the Japan 
Ministry of Education, Culture, Sports,
Science and Technology.


\appendix
\section{Analogy with 3D harmonic oscillator}
\label{a:o3}
MQM in the nontrivial representation has many characters which
are analogous to 3D harmonic oscillator with the Hamiltonian,
\be
\mathcal{H}=-\frac{1}{2}\sum_{i=1}^3 \frac{\partial^2}{\partial x_i^2}
+\frac{1}{2}\sum_i x_i^2=\sum_{i=1}^3
a_i^\dagger a_i+\frac{3}{2}
\ee
with
\be
a_i=\frac{1}{\sqrt{2}}(x_i+ip_i)=\frac{1}{\sqrt{2}}(x_i+\partial_i)\,,\quad
a^\dagger_i=\frac{1}{\sqrt{2}}(x_i-ip_i)=\frac{1}{\sqrt{2}}(x_i-\partial_i)\,.
\ee
In the polar coordinate, the Hamiltonian is rewritten as,
\be
\mathcal{H}=-\frac{1}{2}\left(
\frac{1}{r}\frac{\partial^2}{\partial r^2}r+\frac{1}{r^2}\hat\Omega
\right)+\frac{1}{2}r^2\,,\quad
\hat\Omega=\frac{1}{\sin\theta}\frac{\partial}{\partial\theta}
\left(\sin\theta\frac{\partial}{\partial \theta}\right)
+\frac{1}{\sin^2\theta}\frac{\partial^2}{\partial\phi^2}\,.
\ee
If we write the wave function as $\psi=R(r)Y_{lm}(\theta,\phi)$,
the Schr\"odinger equation for $R(r)$ becomes,
\be
-\frac{1}{2}\left(
\frac{1}{r}\frac{\partial^2(rR)}{\partial r^2}
\right)+\frac{1}{2}\left(
r^2+\frac{l(l+1)}{r^2}
\right)R(r)=ER(r)
\ee
The correspondence between MQM and 3D harmonic oscillator are,
$A^\pm \leftrightarrow   a_i, a_i^\dagger$,
the eigenvalues of $X\leftrightarrow r$,
$U(N)$ rotation $\leftrightarrow SO(3)$,
representation of $U(N)$ $\leftrightarrow$ total angular momentum $l$,
$\sum_{i,j}\frac{\rho(E_{ij})\rho(E_{ji})}{(x_i-x_j)^2}
\leftrightarrow  \frac{l(l+1)}{r^2}$
and so on.

The symmetry of the system is $SO(3)$ rotation generated by
\begin{eqnarray}
 L_i=i\epsilon_{ijk}x_j\partial_k=i\epsilon_{ijk}a^\dagger_i a_j
\end{eqnarray}
which commutes with the Hamiltonian.  There is another $SL(2)$ algebra
which commutes with $L_i$,
\begin{eqnarray}
 Q^-=\frac{1}{2}\sum_i (a_i^\dagger)^2,\quad
 Q^0=\frac{1}{2}(\sum_i a_i^\dagger a_i+\frac{3}{2})=\frac{1}{2}
\mathcal{H},\quad
 Q^+=\frac{1}{2}\sum_i (a_i)^2.
\end{eqnarray}
which satisfy the $sl(2,R)$ algebra,
\begin{equation}
\left[Q^+,Q^-\right]=2Q^0,\quad
\left[Q^0,Q^\pm\right]=\mp Q^\pm\,.
\end{equation}
This is an analogue of $\hW_\infty$.
There is a relation between Casimir operator of $so(3)$ and
$sl(2,R)$ generators,
\begin{eqnarray}
\sum_{i=1}^3 L_i^2=-4Q^-Q^++(2Q^0-\frac{3}{2})(2Q^0-\frac{1}{2})
=-4Q^-Q^++(\mathcal{H}-\frac{3}{2})(\mathcal{H}-\frac{1}{2})
\end{eqnarray}
The shift appearing in $Q^0$ is due to the ground state energy
of the harmonic oscillators.
An analogue of this relation should exist for MQM
which describes the correspondence between the representations
of $U(N)$ and $\hW_\infty$.  So far, since we do not
have a full understanding of the  representation of $\hW_\infty$,
it is difficult to guess such relation.

The Hilbert space of the system is generated
by the direct product of the irreducible representations
of $so(3)$ ($L_i$) and $sl(2,R)$ ($Q^\pm, Q^0$).
For the $so(3)$ algebra, we have spin $l$  representation,
$|l,m\rangle$ ($m=-l,-l+1,\cdots,l-1,l$).
From this state, we generate irrep of
$sl(2,R)$ algebra as
\begin{equation}\label{so3state}
 |n,l,m\rrangle\propto (Q^-)^n |l,m\rrangle\,,
\quad Q^+|l,m\rrangle=0\,,\quad
Q^0|l,m\rrangle=\frac{1}{2}(l+\frac{3}{2})|l,m\rrangle\,.
\end{equation}
We have changed notation $|l,m\rangle\rightarrow |l,m\rrangle$
since they are the ground state of $sl(2,R)$.
The assignment of the weight is necessary since
we have to impose,
\begin{eqnarray}
 L^2|l,m\rrangle=l(l+1)|l,m\rrangle
=(2Q^0-\frac{3}{2})(2Q^0-\frac{1}{2})|l,m\rrangle\,.
\end{eqnarray}
The state $|n,l,m\rrangle$ ($n=0,1,2,\cdots$, $l=0,1,2\cdots$, 
$m=-l,\cdots,l$) span the Hilbert space of the system.

For example, the lower states are given as follows,
\begin{eqnarray}
 \mbox{level $0$}&: &|0\rangle\leftrightarrow|0,0\rrangle\,,\nn\\
 \mbox{level $1$}&: &a_i^\dagger|0\rangle\leftrightarrow
|1,m\rrangle\,,\nn\\
 \mbox{level $2$} &: & a_i^\dagger a_j^\dagger|0\rangle
\leftrightarrow |2,m\rrangle\,,\, Q^-|0,0\rrangle\nn\\
 \mbox{level $3$} &: & a_i^\dagger a_j^\dagger a_k^\dagger|0\rangle
\leftrightarrow |3,m\rrangle\,,\, Q^-|1,m\rrangle
\end{eqnarray}
The partition function associated with spin $l$ have the form,
$(2l+1)q^{l+3/2}\frac{1}{1-q^2}$ where $(2l+1)$ is the number of state
of spin $l$ representation, $l+3/2$ is the ``ground state energy''
for spin $l$ and $\frac{1}{1-q^2}$ is the partition function of 
$sl(2,R)$ representation. The total partition function is
\begin{eqnarray}\label{so3tot}
 \mbox{Tr}q^{\mathcal{H}-3/2}=\sum_{l=0}^\infty
 (2l+1)q^{l}\frac{1}{1-q^2}=\frac{1}{(1-q)^3}\,.
\end{eqnarray}
The right hand side is the partition function of three harmonic
oscillators.

%

\section{Explicit form of the states constructed from Young symmetrizer at the lower levels}
\label{a:lower}
In order to see the relation between the states constructed from
the Young symmetrizer and free boson (fermion) states,
we present the explicit forms of the former at the lower levels.

\paragraph{Level 2}
When $n=2$, the Young symmetrizers are,
$e_{[2]}=\frac{1}{2}(1+(12))$, $e_{[1,1]}=\frac{1}{2}(1-(12))$
and the states that corresponds to them are
\begin{eqnarray}
&& \Psi^{[2],[1^2]}
=\frac{1}{2}\left(\mbox{Tr}(\cC_1 \bar A)\mbox{Tr}(\cC_2 \bar A)
\pm \mbox{Tr}(\cC_1 \bar A\cC_2 \bar A)\right)|0\rangle\,.
\end{eqnarray}
The eigenvalue of $H_2$ is $\pm 2$ for $\Psi^{[2]}$ ($\Psi^{[1^2]}$).
By restricting $\cC_1=\cC_2=1$, $\Psi^{[Y]}$ reduces
to Schur polynomial,
\begin{align}
 \Psi^{[2]} &\rightarrow \frac{1}{2}(J_1^2+J_2)|0\rangle = 
s_{[2]}(J)\ket{0}\,, \quad
 \Psi^{[1^2]} \rightarrow \frac{1}{2}(J_1^2-J_2)|0\rangle = 
s_{[1^2]}(J)\ket{0}\,.
\end{align}
As for the restriction to the adjoint sector, with traceless
matrix $\cC$,
\begin{equation}
 \Psi^{[2]}\rightarrow \frac{1}{2}(J_1\mbox{Tr}(\cC \bar A)+
\mbox{Tr}(\cC\bar A^2))|0\rangle\,,\qquad
 \Psi^{[1^2]}\rightarrow \frac{1}{2}(J_1\mbox{Tr}(\cC \bar A)-
\mbox{Tr}(\cC\bar A^2))|0\rangle
\end{equation}
\paragraph{Level 3}
Young symmetrizers are,
\begin{eqnarray}
 &&e_{[3]}=\frac{1}{3!}(1+(123)+(213)+(12)+(13)+(23))\,,\\
 &&e_{[1^3]}=\frac{1}{3!}(1+(123)+(213)-(12)-(13)-(23))\,,\\
 && e_{[2,1]}=\frac{1}{3}(1+(12)-(13)-(123))\,.
\end{eqnarray}
There are several independent boards associated with $Y=[2,1]$.
Here we pick up the following one:

\begin{center}
\unitlength 0.1in
\begin{picture}(  4,  4)( 30,-20)
\put(31,-17){\makebox(0,0){1}}%
%
\special{pn 8}%
\special{pa 3000 1600}%
\special{pa 3200 1600}%
\special{pa 3200 1800}%
\special{pa 3000 1800}%
\special{pa 3000 1600}%
\special{fp}%
%
\special{pn 8}%
\special{pa 3000 1800}%
\special{pa 3200 1800}%
\special{pa 3200 2000}%
\special{pa 3000 2000}%
\special{pa 3000 1800}%
\special{fp}%
%
\special{pn 8}%
\special{pa 3200 1600}%
\special{pa 3400 1600}%
\special{pa 3400 1800}%
\special{pa 3200 1800}%
\special{pa 3200 1600}%
\special{fp}%
\put(33,-17){\makebox(0,0){2}}%
\put(31,-19){\makebox(0,0){3}}%
\end{picture}
\end{center}
\vspace{-12pt}
From these projectors, one obtains the eigenstates of $H_2$ as,
\begin{eqnarray}
&&  \Psi^{[3]}=\frac{1}{3!}\left(
\mbox{Tr}(\cC_1 \bar A)\mbox{Tr}(\cC_2 \bar A)\mbox{Tr}(\cC_3 \bar A)
+\mbox{Tr}(\cC_1 \bar A\cC_2 \bar A\cC_3\bar A)
+\mbox{Tr}(\cC_2 \bar A\cC_1 \bar A\cC_3\bar A)\right.\nn\\
&&~~~\left.
+\mbox{Tr}(\cC_1 \bar A\cC_2 \bar A)\mbox{Tr}(\cC_3 \bar A)
+\mbox{Tr}(\cC_1 \bar A\cC_3 \bar A)\mbox{Tr}(\cC_2 \bar A)
+\mbox{Tr}(\cC_2 \bar A\cC_3 \bar A)\mbox{Tr}(\cC_1 \bar A)
\right)|0\rangle\,,\nn\\
&&  \Psi^{[1^3]}=\frac{1}{3!}\left(
\mbox{Tr}(\cC_1 \bar A)\mbox{Tr}(\cC_2 \bar A)\mbox{Tr}(\cC_3 \bar A)
+\mbox{Tr}(\cC_1 \bar A\cC_2 \bar A\cC_3\bar A)
+\mbox{Tr}(\cC_2 \bar A\cC_1 \bar A\cC_3\bar A)\right.\nn\\
&&~~~\left.
-\mbox{Tr}(\cC_1 \bar A\cC_2 \bar A)\mbox{Tr}(\cC_3 \bar A)
-\mbox{Tr}(\cC_1 \bar A\cC_3 \bar A)\mbox{Tr}(\cC_2 \bar A)
-\mbox{Tr}(\cC_2 \bar A\cC_3 \bar A)\mbox{Tr}(\cC_1 \bar A)
\right)|0\rangle\,,\nn\\
&&\Psi^{[2,1]}=\frac{1}{3}\left(
\mbox{Tr}(\cC_1 \bar A)\mbox{Tr}(\cC_2 \bar A)\mbox{Tr}(\cC_3 \bar A)
+\mbox{Tr}(\cC_1 \bar A\cC_2 \bar A)\mbox{Tr}(\cC_3 \bar A)\right.\nn\\
&&~~~\left.
-\mbox{Tr}(\cC_1 \bar A\cC_3 \bar A)\mbox{Tr}(\cC_2 \bar A)
-\mbox{Tr}(\cC_1 \bar A\cC_2 \bar A\cC_3\bar A)
\right)|0\rangle\,,
\end{eqnarray}
with
\begin{equation}
 H_2 \Psi^{[3]}=6\Psi^{[3]}\,,\quad
 H_2 \Psi^{[1^3]}=-6\Psi^{[1^3]}\,,\quad
 H_2 \Psi^{[2,1]}=0\,.
\end{equation}
As before, the restriction to the singlet gives
the corresponding Schur polynomials.
The restriction to the adjoint is also similar.
One important lesson here is that there are two
independent states 
which can be derived from
$\Psi^{[2,1]}$.  By putting 
$(\cC_1,\cC_2,\cC_3)=(\cC,1,1)$
and
$(\cC_1,\cC_2,\cC_3)=(1,\cC,1)$
\begin{equation}
 \frac{1}{3}\left(
\mbox{Tr}(\cC\bar A)J_1^2-\mbox{Tr}(\cC \bar A^3)
\right)|0\rangle
\,,\quad
 \frac{1}{3}\left(
\mbox{Tr}(\cC\bar A)J_1^2
+\mbox{Tr}(\cC\bar A^2)J_1
-\mbox{Tr}(\cC\bar A)J_2
-\mbox{Tr}(\cC \bar A^3)\right)|0\rangle\,.
\end{equation}
It explains the degeneracy 2 of the adjoint sector
for $Y=[2,1]$.

\end{document}